\documentclass[preprint]{aastex}
\def\sqd{{deg$^{2}$}}
\def\etal{{\it et al. }}
\def\kms{km~s$^{-1}$~}

\def\msun{$M_\odot$}
\def\mhi{$M_{HI}$}
\def\be{\begin{equation}}
\def\ee{\end{equation}}

\begin{document}
\title{The Arecibo Legacy Fast ALFA Survey: \\
       I. Science Goals, Survey Design and Strategy 
}

\author {Riccardo Giovanelli\altaffilmark{1}, Martha P. Haynes\altaffilmark{1}, 
Brian R. Kent\altaffilmark{1}, Philip Perillat\altaffilmark{2}, Amelie Saintonge\altaffilmark{1},
Noah Brosch\altaffilmark{3}, Barbara Catinella\altaffilmark{2}, G. Lyle Hoffman\altaffilmark{4}, 
Sabrina Stierwalt\altaffilmark{1}, Kristine Spekkens\altaffilmark{1}, Mikael S. Lerner\altaffilmark{2},
Karen L. Masters\altaffilmark{1}, Emmanuel Momjian\altaffilmark{2}, 
Jessica L. Rosenberg\altaffilmark{5}, 
Christopher M. Springob\altaffilmark{1}, Alessandro Boselli\altaffilmark{6}, 
Vassilis Charmandaris\altaffilmark{7}, Jeremy K. Darling\altaffilmark{8}, 
Jonathan Davies\altaffilmark{9}, Diego Garcia Lambas\altaffilmark{10}, 
Giuseppe Gavazzi\altaffilmark{11}, Carlo Giovanardi\altaffilmark{12}, 
Eduardo Hardy\altaffilmark{13}, Leslie K. Hunt\altaffilmark{14}, 
Angela Iovino\altaffilmark{15}, Igor D. Karachentsev\altaffilmark{16}, 
Valentina E. Karachentseva\altaffilmark{17}, Rebecca A. Koopmann\altaffilmark{18}, 
Christian Marinoni\altaffilmark{15}, Robert Minchin\altaffilmark{9}, 
Erik Muller\altaffilmark{19}, Mary Putman\altaffilmark{20},
Carmen Pantoja\altaffilmark{21}, 
John J. Salzer\altaffilmark{22}, Marco Scodeggio\altaffilmark{23}, 
Evan Skillman\altaffilmark{24}, Jose M. Solanes\altaffilmark{25}, 
Carlos Valotto\altaffilmark{10}, Wim van Driel\altaffilmark{26}, 
Liese van Zee\altaffilmark{27}
}

\altaffiltext{1}{Center for Radiophysics and Space Research
and National Astronomy and Ionosphere Center, 
Cornell University, Ithaca, NY 14853. {\it e--mail:} riccardo@astro.cornell.edu,
haynes@astro.cornell.edu, bkent@astro.cornell.edu, amelie@astro.cornell.edu,
sabrina@astro.cornell.edu, spekkens@astro.cornell.edu, masters@astro.cornell.edu,
springob@astro.cornell.edu}

\altaffiltext{2}{Arecibo Observatory, National Astronomy and Ionosphere Center,
Arecibo, PR 00612. {\it e--mail:} bcatinel@naic.edu, lerner@naic.edu, phil@naic.edu, emomjian@naic.edu}

\altaffiltext{3}{The Wise Observatory \& The School of Physics and Astronomy, 
Raymond \& Beverly Sackler Faculty of Exact Sciences, Tel Aviv University, Israel.
{\it e--mail:} noah@wise.tau.ac.il}

\altaffiltext{4}{Hugel Science Center, Lafayette College, Easton, PA 18042.
{\it e--mail:} hoffmang@lafayette.edu}

\altaffiltext{5}{Harvard--Smithsonian Center for Astrophysics, 60 Garden St. MS 65, 
Cambridge MA 02138--1516. {\it e--mail:} jlrosenberg@cfa.harvard.edu}

\altaffiltext{6}{Laboratoire d'Astrophysique, Traverse du Siphon, BP8,
13376 Marseille, France. {\it e--mail:} alessandro.boselli@oamp.fr}

\altaffiltext{7}{Dept. of Physics, University of Crete, 71003 Heraklion, Greece.
{\it e--mail:} vassilis@physics.uoc.gr}

\altaffiltext{8}{Carnegie Observatories, 813 Sta. Barbara St., Pasadena, CA 91101.
{\it e--mail:} darling@ociw.edu}

\altaffiltext{9}{Dept. of Physics \& Astronomy, U. of Wales, Cardiff
CF243YB, United Kingdom. {\it e--mail:}
jonathan.davies@astro.cf.ac.uk, robert.minchin@astro.cf.ac.uk}

\altaffiltext{10}{Observatorio Astronomico, U. de Cordoba, Cordoba 5000, Argentina.
{\it e--mail:} dgl@oac.uncor.edu, val@oac.uncor.edu}

\altaffiltext{11}{Dept. of Physics, Univ. di Milano--Bicocca, Milano 20126, Italy.
{\it e--mail:} giuseppe.gavazzi@mib.infn.it}

\altaffiltext{12}{INAF, Osservatorio Astrofisico d'Arcetri, Largo Enrico
Fermi 5, 50125 Firenze, Italy. {\it e--mail:} giova@@arcetri.astro.it}

\altaffiltext{13}{National Radio Astronomy Observatory, Apoquindo
3650, Piso 18, Las Condes, Santiago, Chile. {\it e--mail:} ehardy@nrao.edu}

\altaffiltext{14}{INAF, Istituto di Radioastronomia/Sez. Firenze, Largo Enrico
Fermi 5, 50125 Firenze, Italy. {\it e--mail:} hunt@arcetri.astro.it}

\altaffiltext{15}{Osservatorio Astronomico di Brera, INAF, Via Brera 28, 
20121 Milano, Italy. {\it e--mail:} iovino@brera.mi.astro.it, marinoni@brera.mi.astro.it}

\altaffiltext{16}{Special Astrophysical Observatory, Russian Academy of Sciences, 
Niszhnij Arkhyz 369167, Zelencukskaya, KChR, Russia. {\it e--mail}: ikar@sao.ru}

\altaffiltext{17}{Dept. of Astronomy \& Space Science, Kyiv University, Kyiv 
252017, Ukraine. {\it e--mail:} vkarach@observ.univ.kiev.ua} 

\altaffiltext{18}{Dept. of Physics \& Astronomy, Union College, Schenectady, NY 12308.
{\it e--mail:} koopmanr@union.edu}

\altaffiltext{19}{ATNF, CSIRO, PO Box 76, Epping, NSW 1710, Australia; {\it e--mail:} 
erik.muller@atnf.csiro.au} 

\altaffiltext{20}{Astronomy Dept., U. of Michigan, Ann Arbor, MI 48109.
{\it e--mail:} mputman@umich.edu}

\altaffiltext{21}{Dept. of Physics, U. of Puerto Rico. P.O. Box 364984, San Juan, 
PR 00936. {\it e--mail:} cpantoja@naic.edu}

\altaffiltext{22}{Astronomy Dept., Wesleyan University, Middletown, CT 06457.
{\it e--mail:} slaz@astro.wesleyan.edu}

\altaffiltext{23}{Dipartimento di Fisica, U. di Milano, Via Celoria 16, 20133 
Milano, Italy. {\it e--mail:} marcos@mi.iasf.cnr.it}

\altaffiltext{24}{Astronomy Dept., U. of Minnesota, 116 Church St. SE, Minneapolis, 
MN 55455. {\it e--mail:} skillman@astro.umn.edu}

\altaffiltext{25}{Department d'Astronomia i Meteorologia, U. de Barcelona, Av.
Diagonal 647, 08028 Barcelona, Spain, and Centre Especial de Recerca en Astrof\'\i sica,
F\'\i sica de Part\'\i cules i Cosmologia associated with the Instituto de Ciencias
del Espacio, Consejo Superior de Investigaciones Cient\'\i ficas. {\it e-mail:} 
jm.solanes@ub.edu}

\altaffiltext{26}{Observatoire de Meudon, 5 Place Jules Janssen, 92195 Meudon, France.
{\it e--mail:} wim.vandriel@obspm.fr}

\altaffiltext{27}{Astronomy Dept., Indiana University, Bloomington, IN  47405. 
vanzee@astro.indiana.edu}

\begin{abstract}
The recently initiated Arecibo Legacy Fast ALFA (ALFALFA) survey aims to map 
$\sim$ 7000 \sqd ~of the high galactic latitude sky visible from Arecibo, providing
a HI line spectral database covering the redshift range between -1600 \kms and 18,000 \kms 
with $\sim 5$ \kms ~resolution. Exploiting Arecibo's large collecting area and small
beam size, ALFALFA is specifically designed to probe the faint end of the HI mass 
function in the local universe and will provide a census of HI in 
the surveyed sky area to faint flux limits, making it especially useful in synergy 
with wide area surveys conducted at other wavelengths.
ALFALFA will also provide the basis for studies of the dynamics
of galaxies within the Local and nearby superclusters, will allow measurement of the
HI diameter function, and enable a first wide-area blind search for local HI tidal
features, HI absorbers at $z < 0.06$ and OH megamasers in the redshift range
$0.16 < z < 0.25$. Although completion of the survey will require some
five years, public access to the ALFALFA data and data products will be provided
in a timely manner, thus allowing its application for studies beyond those 
targeted by the ALFALFA collaboration. ALFALFA adopts a two-pass, minimum 
intrusion, drift scan observing technique which samples the same region of sky
at two separate epochs to aid in the discrimination of cosmic signals from
noise and terrestrial interference. Survey simulations, which take into account 
large scale structure in the mass distribution and incorporate experience with the 
ALFA system gained from tests conducted during its commissioning phase, suggest that 
ALFALFA will detect on the order of 20,000 extragalactic HI line
sources out to $z \sim 0.06$, including several hundred with HI masses 
\mhi ~$< 10^{7.5}$ \msun.

\end{abstract}


\section {Introduction}

The first 21 cm line detection of an extragalactic source (the Magellanic 
Clouds) was achieved by Kerr \& Hindman (1953) with a 36--foot transit 
telescope just over half a century ago. The construction of large, single 
dish radio telescopes produced seminal discoveries in the decade of the 1960s, as
illustrated in the fundamental paper of that period (Roberts 1975). A 
decade later, the completion of the Very Large Array (VLA) and of the Westerbork
Synthesis Radio Telescope, the resurfacing of the Arecibo dish and 
rapid progress in detector and spectrometer technology made it possible 
for HI spectroscopy to achieve order of magnitude improvements in terms of 
sensitivity and resolution. New scientific problems became accessible, and 
extragalactic HI line research underwent a phase of rapid growth. The study of rotation 
curves led to the discovery of dark matter in spiral galaxies; the
potential of the luminosity--linewidth relation as a cosmological tool 
became apparent; the impact of tidal interactions 
and of the intracluster medium on galaxy evolution was illustrated in great 
detail through measures of the HI emission. The 21 cm line was found
to be an expedient tool to determine accurate galaxy redshifts,
playing an important role in confirming the filamentary nature of the 
large--scale structure of the Universe. The application of the luminosity--linewidth
relation led to accurate estimates of cosmological parameters 
and to the characterization of the peculiar velocity field in the local
Universe. Highly sensitive measurements in the peripheries of disk galaxies 
revealed edges in their visible components, and a number of optically inert 
objects was discovered. 

Until a few years ago, however, comprehensive wide angle surveys of the extragalactic
HI sky were unavailable. At the close of the last decade, the advent 
of multifeed front--end systems at L--band finally made possible the efficient 
coverage of large sections of the extragalactic sky. The first such system
to be used for that purpose was installed on the 64~m Parkes telescope
in Australia, and has produced the excellent results of the HIPASS
survey (Barnes \etal ~2001; Meyer \etal ~2004). A second 4-feed system on the 76~m Lovell
Telescope at Jodrell Bank produced the HIJASS (Lang \etal ~2003) survey.
The 1990s upgrade of the Arecibo telescope, which
replaced its line feeds with a Gregorian subreflector system, made it possible
for that telescope to host feed arrays, as proposed by Kildal \etal ~(1993).
Eventually built and installed at Arecibo in 2004, this 7-beam radio
``camera'',  named ALFA (Arecibo L--band Feed Array), is now operational,
enabling large--scale mapping projects with the great sensitivity of the 
305--m telescope. A diverse set of mapping projects are now underway, ranging
from extragalactic HI line, to Galactic line and continuum, to pulsar 
searches. Here, we introduce one of these newly-initiated surveys,
specifically designed to map approximately
one fifth of the sky in the HI line, out to a distance of 250 Mpc. 
The survey, currently underway at Arecibo, is referred to as ALFALFA,
the Arecibo Legacy Fast ALFA Survey. 

\begin{figure}[ht]
\epsscale{1.3}
\hskip -6cm
\plotone{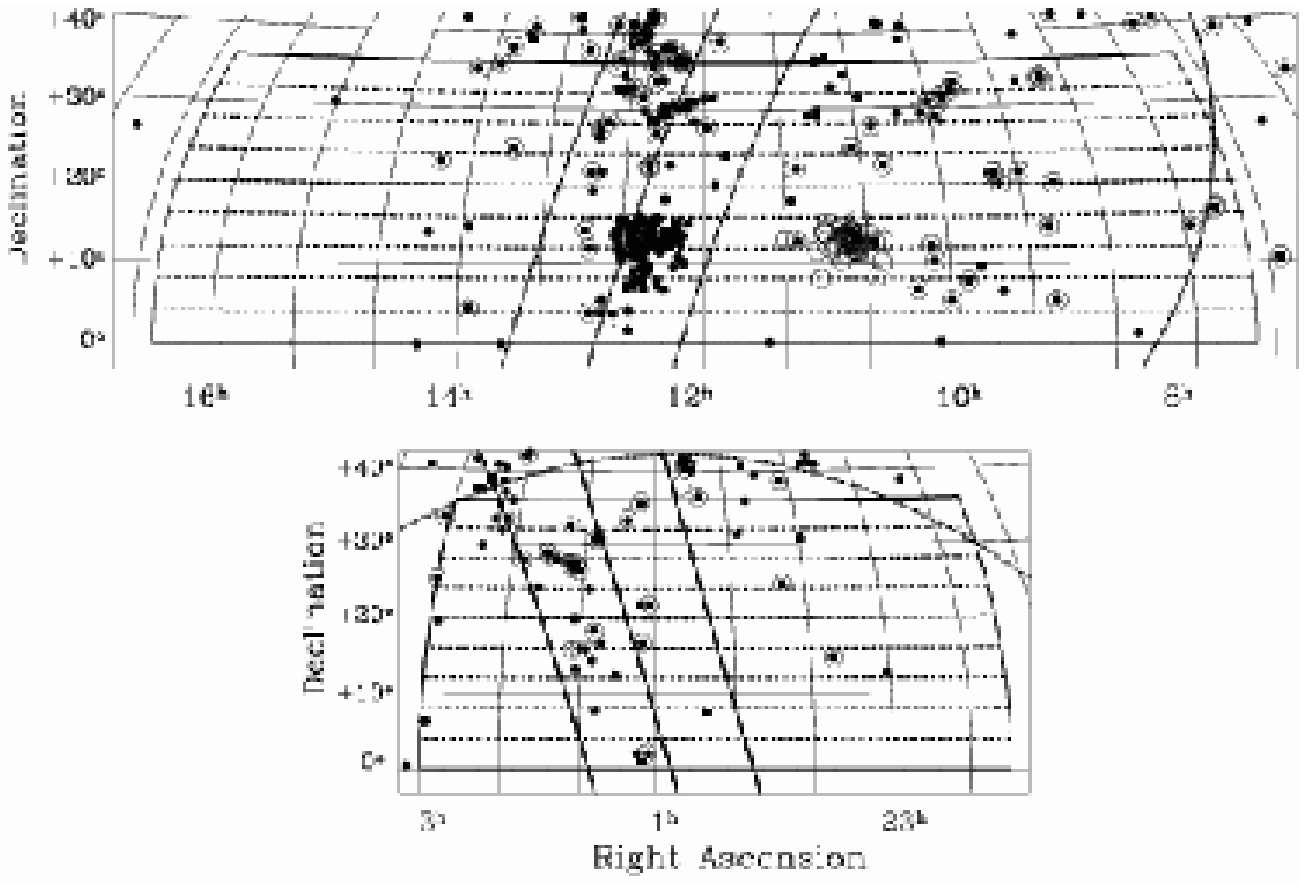}
\vskip -12cm
\caption{Proposed sky coverage of the ALFALFA survey, in the Virgo (upper) and anti-Virgo 
(lower) directions. In each panel, the thicker lines at constant RA or Dec. outline the 
proposed survey area. Dashed lines at constant Dec. make the designated ALFALFA `tile' 
strip boundaries. The thick dotted curves to the right of the upper panel and top of
the bottom panel mark b = $+20^\circ$ (upper) and $-20^\circ$ (lower) while the set of
three thick lines crossing each panel top to bottom trace SGL = $-10^\circ$, $0^\circ$ 
and $+10^\circ$. Filed circles mark galaxies 
with observed heliocentric recessional velocities cz $<$ 700 \kms, while open  
circles denote objects believed to lie within 10 Mpc (Karachentsev \etal ~2004), 
based largely on primary distances.}
\label{skycov}
\end{figure}

As illustrated in Figure \ref{skycov},
ALFALFA aims to cover 7074 \sqd ~of the high galactic latitude sky
between $0^\circ$ and $36^\circ$ in Declination, requiring a total of 4130 
hours of telescope time. Exploiting the large collecting area of the
Arecibo antenna and its relatively small beam size ($\sim
3.5$\arcmin), ALFALFA will be nearly eight times more sensitive than
HIPASS with $\sim$four times better angular resolution. Furthermore,
its spectral backend provides 3 times better spectral resolution 
(5.3 \kms ~at $z = 0$) over 1.4 times more bandwidth. These
advantages, in combination with a simple observing technique designed
to yield excellent baseline characteristics, flux calibration and
HI signal verification, offer new opportunities to explore the 
extragalactic HI sky. Data taking for ALFALFA was initiated in 
February 2005, and, in the practical context of time allocation 
at a widely used, multidisciplinary national facility like Arecibo, 
completion of the full survey is projected to require 5--6 years. 

As discussed in Section \ref{simul}, simulations predict that
ALFALFA will detect some 20,000 extragalactic HI line sources, from
very nearby low mass dwarfs to massive spirals at $z \sim 0.06$.
The survey is designed specifically to determine robustly 
the faint end of the HI mass function (HIMF) in the local universe 
at masses $M_{HI} < 10^8$ $M_\odot$, and will at the same time
provide a census of HI in the surveyed sky area, making it especially useful 
in synergy with other wide area surveys such as SDSS, 2MASS, GALEX, ASTRO--F, etc. 
In conjunction with optical studies of comparable volumes, ALFALFA
will help determine the true census of low mass satellites and the
widely distributed dwarf galaxy population in the Local and
surrounding groups. Its dataset will also provide the basis for studies of the dynamics
of galaxies within the Local and nearby superclusters, will allow measurement of the 
HI diameter function, and will enable a first wide--area blind search for local HI tidal
features, HI absorbers at $z < 0.06$ and OH megamasers in the redshift range 
$0.16 < z < 0.25$. Survey details and status can be found by visiting
its website\footnote{\it http://egg.astro.cornell.edu/alfalfa}.

Survey efforts of this scale and scope require careful optimization
of their operational strategy towards achieving the science
objectives within the constraints imposed by practical observing
conditions and requirements. In this paper, we introduce the science objectives of
ALFALFA, the principal constraints which set its strategy and the
results of survey simulations which allow prediction of its eventual
results. In a companion paper (Giovanelli \etal ~2005; Paper II), we
present results obtained during a precursor observing run, designed to
allow us to test and optimize the ALFALFA strategy during the ALFA
commissioning phase in fall 2004.

We summarize, in 
Section \ref{science}, the main scientific motivations of the survey.
Technical details of the hardware are given in Section \ref{alfa}, while
criteria leading to the design of the survey, in the form of scaling laws and
survey simulations are described in Section \ref{strategy}. Observing modes,
sky tiling and data processing plans are presented in Section \ref{tiles}, while
Section \ref{sensitivity} summarizes sensitivity numbers at various stages of
the survey. We elaborate on the treatment of candidate detections and follow--up
observations in Section \ref{followup} and summarize in Section \ref{summary}. 
Throughout the paper, we assume $H_\circ=70$ \kms Mpc$^{-1}$.

\section{Overview of ALFALFA Science Goals}\label{science}

Following on the results of a number of important and sucessful previous
blind HI surveys, extragalactic HI surveys with ALFA will exploit Arecibo's 
huge collecting area to explore larger volumes of the universe with greater 
sensitivity and higher angular and spectral resolution if they are to break 
new science ground. With Arecibo's tremendous sensitivity and beam size 
advantages, ALFALFA is designed for wide areal coverage, thereby increasing 
the volume sampled locally, yielding a deep, precise census of HI in the 
local Universe to the lowest HI masses. 

ALFALFA aims to survey 7074 \sqd ~of sky at high galactic latitudes which lie 
within the declination limits of the Arecibo telescope, $-2^\circ <$
Decl. $< +38^\circ$, as illustrated in Figure \ref{skycov}. 
Based on simulations described in Section \ref{simul} and now verified by the results of the 
ALFALFA precursor observations presented in Paper II, ALFALFA is 
expected to yield on the order of 20,000 HI line detections, sampling a wide range of 
sources from local, very low HI mass dwarfs to gas-rich massive galaxies 
seen to $z \sim0.06$ ($\sim$250 Mpc). HI spectra provide redshifts, HI masses and rotational
widths for normal galaxies, trace the history of tidal events with high
kinematical accuracy and provide 
quantitative measures of the potential for future star formation via 
comparative HI contents. As a blind HI survey, ALFALFA will not be biased
towards the high surface brightness galaxies typically found in optical 
galaxy catalogs and moreover, in contrast to HIPASS and HIJASS, will have adequate 
angular and spectral resolution to be used on its own, generally without 
the need for follow--up observations to determine identifications, positions 
and, in many cases, HI sizes. The wide areal coverage of ALFALFA overlaps with several
other major surveys, most notably the Sloan Digital Sky Survey (SDSS), 2MASS and
the NVSS. The catalog products of ALFALFA will be invaluable for multiwavelength
data mining for a wide spectrum of purposes, and a key element of our
overall collaborative program is to provide 
broad application, legacy data products that will maximize the science
fallout of the ALFALFA survey.

A primary objective of ALFALFA is the robust determination of the
faint end of the HI mass function (HIMF). The HIMF is the cosmic number 
density, per bin of HI mass, of detectable HI line signals in a survey 
sensitive to the global neutral hydrogen within a system. The most recent 
estimates of the HIMF based on significant numbers of galaxies have been 
presented by Zwaan \etal ~(1997, hereafter Z97), Rosenberg \& Schneider (2002; 
hereafter RS02), Zwaan \etal ~(2004, hereafter Z04), 
Zwaan \etal ~(2005, hereafter Z05) and Springob \etal ~(2005a). 
The Z04 and Z05 HIMFs are based on the HIPASS survey, while the RS02 and 
Z97 HIMFs are both 
based on surveys conducted at Arecibo during the period of its recent
upgrade. The faint end slope of those determinations of the HIMF vary 
between $-1.20$ and $-1.53$, yielding extrapolations below \mhi ~$= 10^7$
\msun~ that disagree by an order of magnitude near $10^6$ \msun, the RS02 
HIMF having the steeper slope.  All of the previous HI blind surveys sample 
a lower mass limit just below \mhi ~$= 10^8$ \msun. No extragalactic HI 
sources were detected by RS02 or Z97 with $M_{HI} < 10^7$ \msun,
while 3 are claimed by Z04, and only a small number of detections have
\mhi ~$< 10^8$ \msun. We note that the distances of those detections are 
highly uncertain, for they are very nearby and the impact of peculiar
velocity on the observed redshift is quite large, as pointed out by
Masters \etal ~(2004). Thus current inferences on the behavior of the HIMF
at low mass levels are quite unreliable, as they are based on very few 
objects of highly uncertain distance. 

With the aim of exploring the HIMF at masses \mhi ~$< 10^8$ \msun,
ALFALFA will cover a very large solid angle in order to survey an
adequate volume at D $<$ 20 Mpc, a distance within which the low HI mass
systems are detectable. As shown by the simulations described in 
Section \ref{simul}, ALFALFA 
will detect several hundred objects with \mhi ~$< 10^{7.5}$ \msun. In
addition, its extensive catalog of more massive objects will allow
comparison of the high mass end, \mhi ~$> 10^9$ \msun, in the 
diverse range of environments found in the volume out to 250 Mpc.
The Arecibo sky to be surveyed by ALFALFA, as shown in Figure \ref{skycov}
includes the rich central regions of the Local Supercluster and the
nearby low density anti-Virgo region as well as a number of more
distant large scale features, most notably the main ridge of the Pisces-Perseus
Supercluster and the Great Wall connecting the Abell 1367 -- Coma and
Hercules superclusters. 

At the Virgo distance, ALFALFA should detect
galaxies with HI masses as low as \msun ~$\sim 10^7$ \msun. 
ALFALFA will cover more than one thousand \sqd~ around Virgo,
yielding a database of unprecedented breadth for combination
with the SDSS, GALEX and other surveys to construct a complete
census of baryon bearing objects in the cluster and its
full infall region. The combination of HI content, HI distribution and
the derived kinematical information with other multiwavelength studies will enable
detailed modelling of the relative efficiency of gas stripping
mechanisms such as tides, ram pressure or galaxy harassment as the
origin of gas deficiency in Virgo. ALFALFA's HI maps will trace
intriguing HI features like the Virgo ``dark cloud'' (Davies \etal ~2004;
Minchin \etal ~2005), the HI ``plume'' around NGC~4388 (Oosterloo \&
van Gorkom 2005) and the huge envelope surrounding NGC~4532 and DDO137
(Hoffman \etal ~1992). In
more quiescent regions than Virgo, extensive tidal
features such as the Leo Triplet (Haynes, Giovanelli \& Roberts 1979),
and enigmatic systems such as the 200~kpc ``Leo ring'' (Schneider \etal ~1983)
may be found. ALFALFA will enable the first truly blind survey for HI tidal 
remnants with both sufficient angular resolution and wide areal coverage to 
verify their nature.

While HI appendages uncover past disruptive events in galaxy evolution, 
extended gas disks around galaxies represent a reservoir for future star 
formation activity. In contrast to HIPASS and HIJASS
which were limited by much poorer angular resolution (15.5\arcmin ~and
12\arcmin, respectively), the 3.5\arcmin 
~beam of ALFA will resolve the HI disks of $\sim$500 gas-rich
galaxies, allowing a quantitative measure of their HI sizes (Hewitt \etal ~1984) and 
the derivation of the HI diameter function. In combination with optical photometry,
ALFALFA will determine the fraction of galaxies with extended gas disks and enable studies 
of their host galaxies, their environments, morphologies and the role of gas in their
evolution. More extremely extended gas disks, such as those found in DDO~154 
(Krumm \& Burstein 1984), UGC~5288 (van Zee 2004) and
NGC~3741 (Begum \etal ~2005) may lurk yet unidentified.
Because of its wide sky coverage, ALFALFA will trace important high-velocity 
cloud (HVC) structures in and around the Milky Way, such as the northern 
portions of the Magellanic Stream and Complex C at several times better
spatial and spectral resolution than HIPASS, particularly important
advantages in the case of narrow linewidth HVC cores (Giovanelli \& Brown 1973).  
Because of its high flux sensitivity, ALFALFA will be eight times
more sensitive than HIPASS to unresolved small clouds, or ultra-compact HVCs. 
While Arecibo cannot reach as far north as M31, ALFALFA will cover part of the 
region containing the clouds in its periphery
identified by Thilker \etal ~(2004) and their possible extension toward the region
around M33 (Westmeier, Braun \& Thilker 2005). 

In addition to the study of HI in emission, ALFALFA will provide a
dataset well-suited for a blind survey of HI absorption out to $z \sim
0.06$. The background continuum source counts in the ALFALFA survey
region at 1.4 GHz yield over 2000 sources brighter than 0.4 Jy and
more than 10000 brighter than 0.1 Jy. 
The major practical difficulty with HI line absorption studies
is spectral baseline determination in the presence of standing
waves. The large number of continuum sources present in the ALFALFA 
dataset and the adopted  ``drift'' technique (Section \ref{drift}) will
aid in the assessment of whether a given spectral feature is real absorption. 

By its combination of studies of HI emission and absorption in the local
universe, ALFALFA will allow a robust estimate of the local HI cross 
section, as well as a measure of its clustering correlation amplitude and scale.

In addition to HI line studies, the frequency range of the ALFALFA
survey will also include, serendipitously, lines from OH Megamasers (OHM)
arising from the nuclear molecular regions in merging galaxy systems.
Approximately 100 such sources are known to date, half of which were discovered
recently at Arecibo (e.g. Darling \& Giovanelli 2002). Observations of 
OHMs hold the potential for tracing the merger history of the Universe since the 
sources are associated with merging galaxies. An essential tool in this exercise
is the OHM luminosity funtion at low $z$. ALFALFA should detect several
additional dozen OHMs in the redshift interval 0.16--0.25, and allow a more
robust determination of the low $z$ OHM luminosity function than 
currently available.

\section{ALFA: The Arecibo L--Band Feed Array and its Spectral Line Backend}\label{alfa}

\begin{figure}[h]
\epsscale{0.9}
\vskip 0cm
\hskip -5cm
\plotone{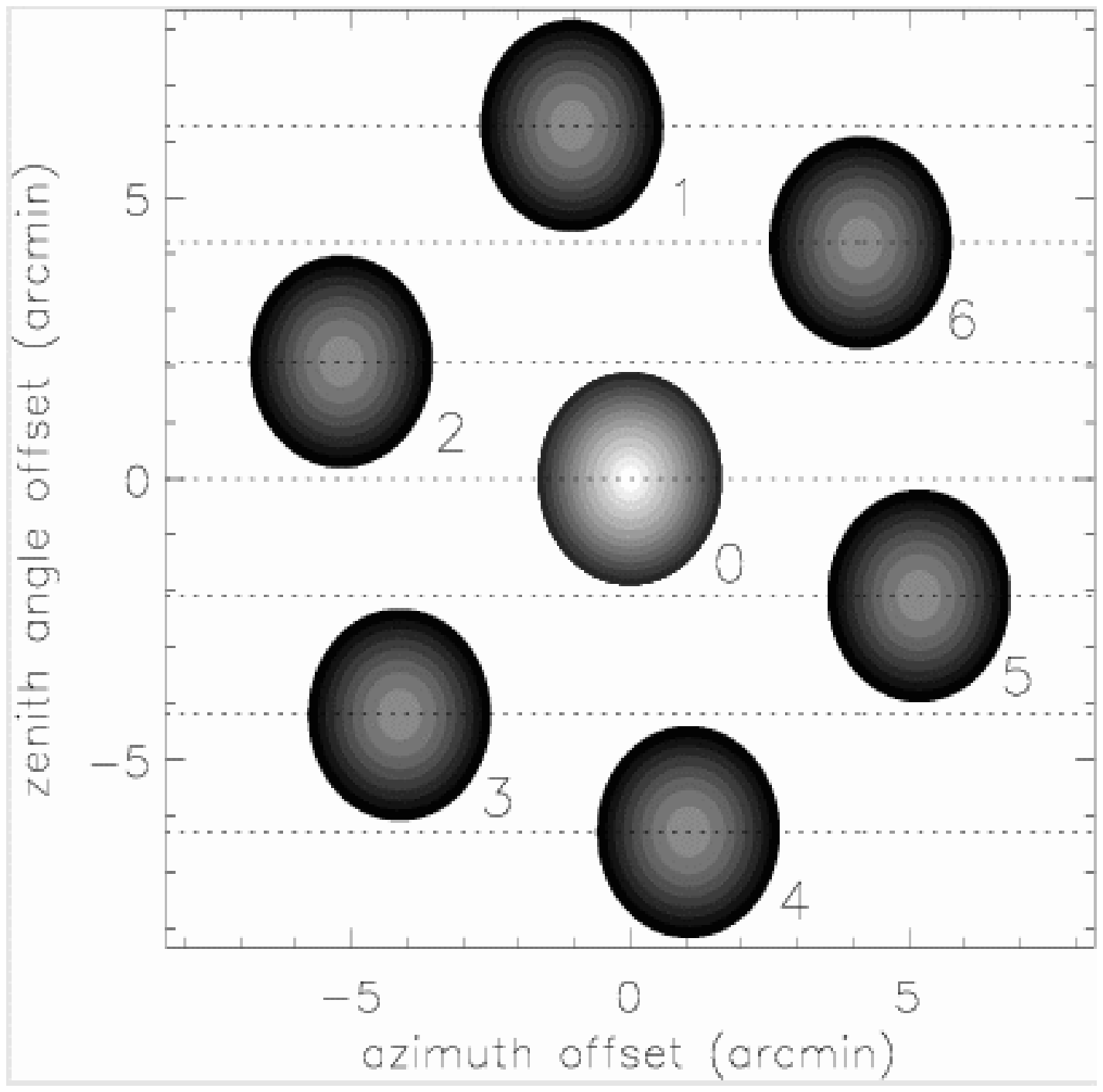}
\vskip -2in
\caption{Sketch of the geometry of the ALFA footprint, with the
array located along the local meridian and rotated by an angle of 
$19^\circ$ about its axis. The outer boundary of each beam corresponds 
to the -3 dB level. The dashed horizontal lines represent the tracks 
at constant Declination of the seven ALFA beams, as data is acquired 
in drift mode.}
\label{alfabeams}
\end{figure}

The construction of the Gregorian subreflector system for the Arecibo telescope,
completed in the late 1990s, made possible the development of focal plane
feed arrays (effectively, {\it creating} a focal plane). This development was
foreseen during the planning phases of the Gregorian upgrade (Kildal \etal 1993).
A seven feed array was commissioned at the Observatory during 2004.
Six of the seven feeds (numbered 1 through 6) are physically arranged on the
corners of a regular hexagon, while the seventh (feed 0) is at its center, as 
shown in Figure \ref{alfabeams}. The feeds
can receive dual, linear polarizations and their spectral response is optimized
for the range 1225--1525 MHz. They are stepped TE$_{11}$ mode horns of
25 cm aperture, as described in Cort\' es--Medell\'\i n (2002). Because 
the optical design of the Gregorian subreflectors maximizes the illuminated
area of the primary by sacrificing its circular symmetry (the illuminated area
is elliptical), a circular pattern in the sky maps on the focal plane as an 
ellipse of axial ratio 1.15; reciprocally, the footprint of the centers of the
outer beams of the ALFA array on the sky is that of a hexagon inscribed in 
an ellipse of that axial ratio. Similarly, the seven beams have an elliptical 
shape of the same axial ratio and orientation as the array pattern. The major
axis of the ellipse is linked to the azimuth of the receiver, so its orientation
on the sky changes with telescope configuration. In Figure \ref{alfabeams}
the relative location of the beams is shown when the array is positioned
at the meridian and rotated about its symmetry axis by $19^\circ$. In this
sketch, the outlines of the beams are shown at the half--power response, 
for which the beam sizes are 3.3\arcmin ~along the azimuth direction and 
3.8\arcmin ~along the zenith angle directon, with small variations from one 
beam to the other. The central beam 0 has higher gain ($\simeq 11$ K/Jy) 
than the peripheral beams 1--6 (gain of $\simeq 8.5$ K/Jy), which is illustrated
in the sketch by the brighter contours. The dotted lines indicate the tracks of
constant Declination made by each of the beams, when data is acquired in
drift mode. Projected on the sky, the ALFA footprint in this configuration
is such that beam 1 points farthest to the North and beam 2 farthest to
the West, for observations South of the Zenith. For observations North
of the Zenith, beam 1 points farthest to the South and beam 2 farthest
to the East.

\begin{figure}[h]
\plotone{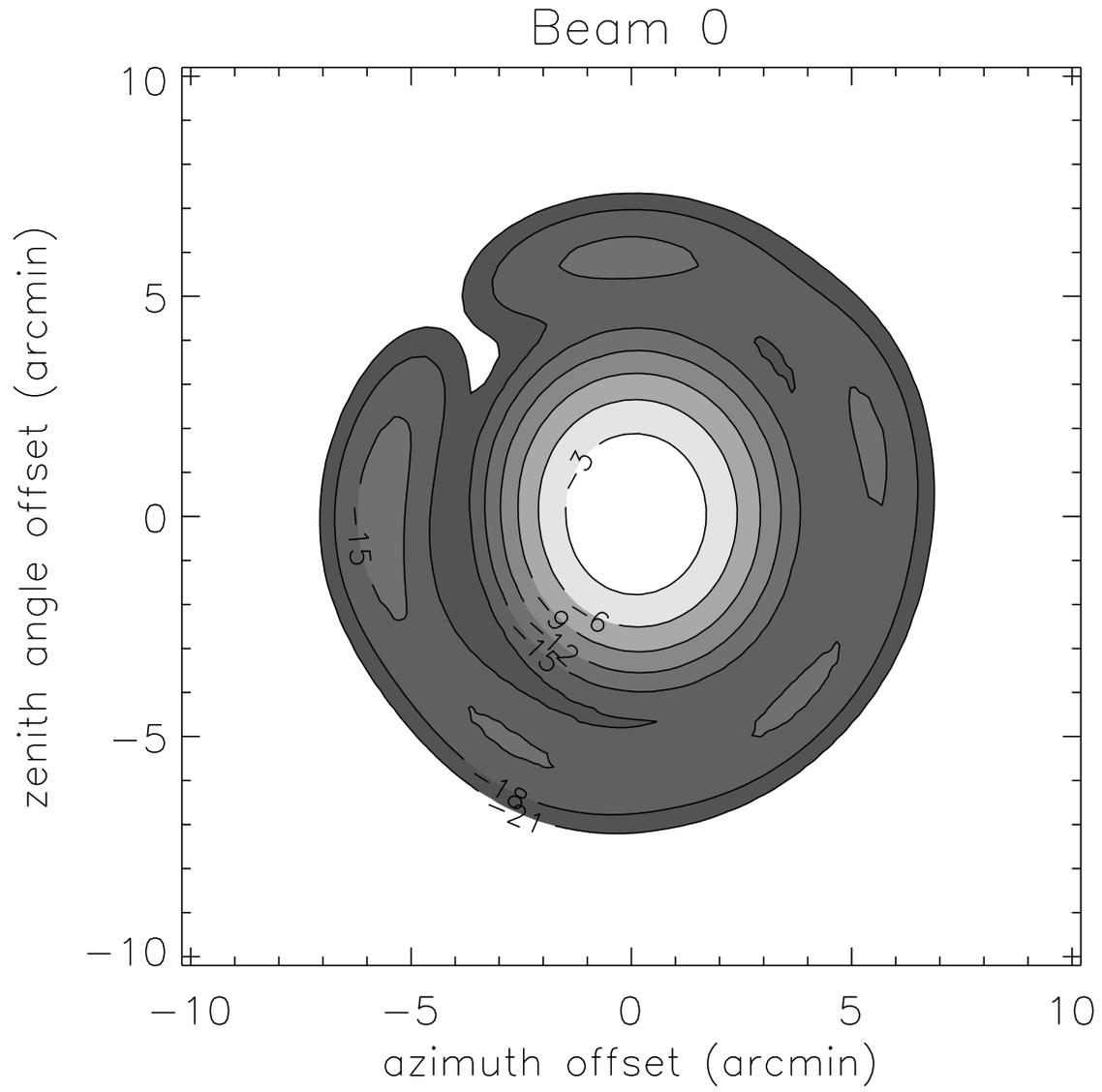}
\caption{Beam pattern of beam 0. Contour lines and shading intervals 
are plotted at intervals 
of 3 dB below peak response (the highest contour is at half the peak power). 
The first sidelobe ring, with a diameter near 12', is at approximately
-15 dB.}
\label{sidelobe0}
\end{figure}

\begin{figure}[h]
\hskip -3cm
\plotone{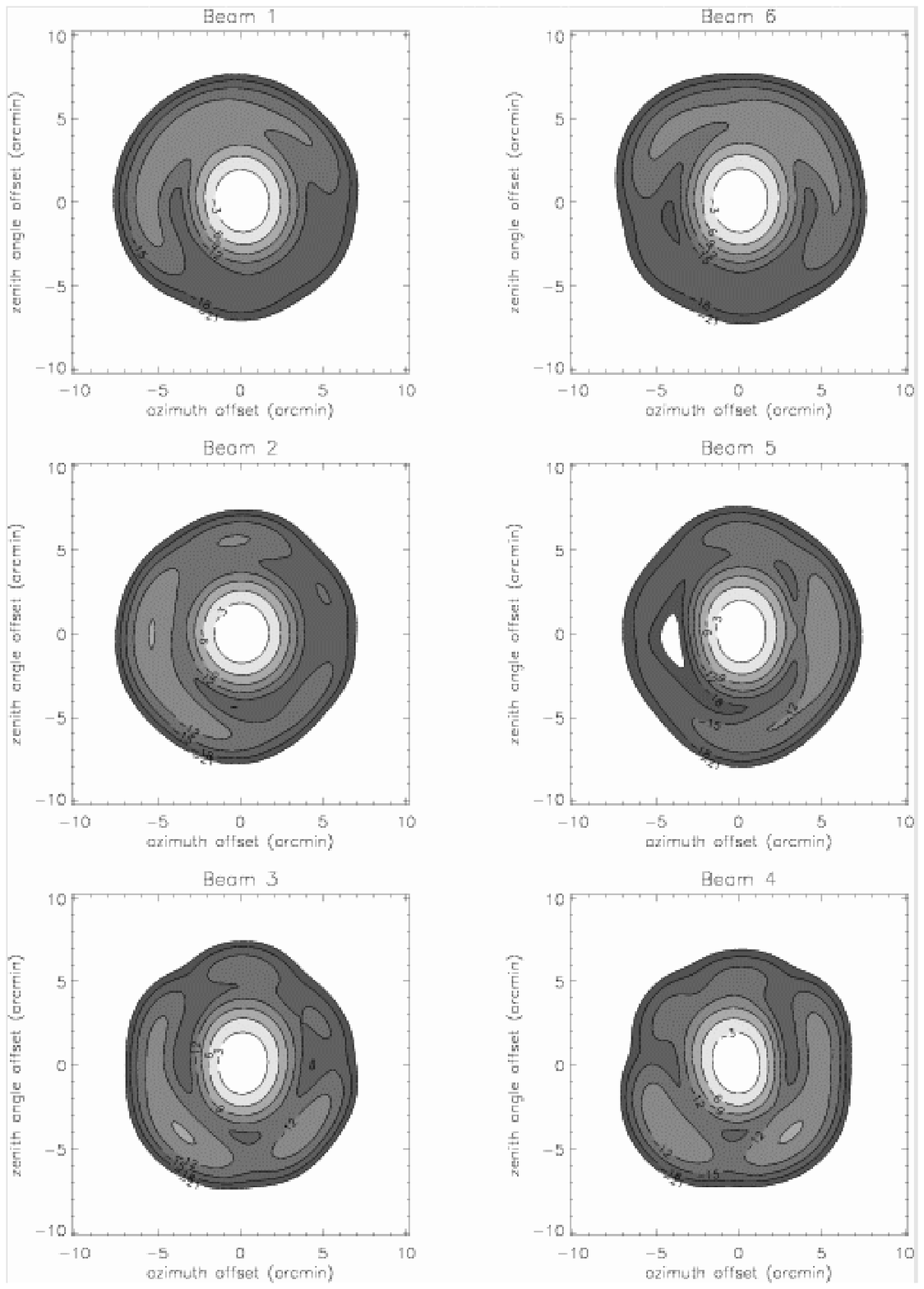}
\vskip -5cm
\caption{Beam patterns of the six peripheral ALFA beams. Contour lines 
and shading intervals are 
plotted at intervals of 3 dB below peak response (the highest contour is 
at half the peak power). Note that the sidelobe levels are significantly 
larger than for the central beam 0, and that they rise steeply on the outer 
side of the array, exhibiting strong comatic aberration.}
\label{sidelobe6}
\end{figure}

Figures \ref{sidelobe0} ~and \ref{sidelobe6}~ show the pattern for each
of the ALFA beams, obtained by mapping the radio source 3C 138 near
transit. Sidelobe levels are very different for each of the beams of ALFA. 
Located at the center of the array, beam 0 has the most symmetric beam pattern,
with a first sidelobe ring near 15 dB below the response at beam center,
as shown in Figure \ref{sidelobe0}. Contour levels are plotted at intervals
of 3 dB. The outer beams have a very marked comatic aberration, as shown 
in Figure \ref{sidelobe6}. The first sidelobe 
ring of the outer beams is strongly asymmetric, reaching levels near 7--8 dB 
below peak response, on the section away from the array center. This feature 
of the system will require careful attention, especially in the analysis of 
data obtained in the vicinity of strong and/or extended sources. 

The system
temperature ranges between 26 and 30 K for all beam/polarization
channels, when pointing away from strong continuum sources.

The array can be rotated about its axis, centered on beam 0, and thus the
relative position of the beams on the sky can be rotated along the 
elliptical perimeter. In the case of drift mode observations, it is desirable 
to position the array in such a manner that the beam tracks are equally 
spaced in Declination. Because of the ellipticity of the array pattern on
the sky, the separation between beam tracks depends on both the
array rotation angle as well as on the array azimuth. When the telescope 
feed arm is stationed along the local meridian, the optimal array rotation 
angle is $19^\circ$, as shown in Figure \ref{alfabeams}. In that case, beam 
tracks are spaced 2.1\arcmin ~in Declination. A single drift will thus sweep 
seven nearly equidistant tracks covering 14.5\arcmin ~in Declination, at 
slightly below the Nyquist sampling rate. ALFALFA will map most of the 
extragalactic sky in drift mode with
ALFA stationed along the local meridian, at a local azimuth of either
$0^\circ$ (for observations South of Zenith) or $180^\circ$ (for North
of Zenith). Only Declination tracks transiting within $2^\circ$
of the Zenith will be mapped with ALFA at an azimuth near $90^\circ$ or
$270^\circ$ in order to avoid impractically small zenith angles. 
The beam separation for equidistant tracks is 1.8\arcmin ~at these
azimuths, and thus the Declination sampling will be denser,
as the elliptical pattern of the sky footprint of both the array and the
individual beams will have its major axis oriented nearly parallel to
the drift direction. Our survey will thus be done with ALFA in only two
sets of configurations: one for all observations between Declinations 
$0^\circ$ and $16^\circ$, as well as between $20^\circ$ and $36^\circ$, with
ALFA on the meridian, and a second for observations between Declinations
$16^\circ$ and $20^\circ$.

Spectra will be recorded every second, yielding approximately 14 samples 
per beam in the Right Ascension direction. This sampling rate, which 
largely exceeds Nyquist, is principally motivated by the advantages 
deriving in the identification of radio frequency interference (RFI).
Further details on ALFA can be found at the NAIC website
\footnote{\it http://alfa.naic.edu}.

ALFALFA uses the feed array connected through a fiber optics IF--LO
system to a spectral line, digital backend consisting of a set of processors 
each individually referred to as a WAPP (Wideband Arecibo
Pulsar Processor). The full spectral backend consists of four WAPP units,
each capable of processing the two polarization signals from two ALFA beams.
The WAPP set can thus produce 16 autocorrelation spectra, each with a 
maximum bandwidth of 100 MHz, over 4096 lags.
Fourteen of those are matched 
to the seven polarization pairs from the ALFA beams, and a spare pair 
duplicates the signal of the seventh beam. At the offline processing stage, 
the extra pair of spectra are used for RFI monitoring purposes. Each data
record thus consists of 65,536 spectral samples (16 $\times$
4096). Since, as mentioned above, 
ALFALFA records data every second, the generation of raw data by the
survey is slightly over 1 GB per hour, including headers.

\section{Survey Design}\label{strategy}

The strategy for the ALFALFA survey has been developed over the last few years,
balancing the practical realities involved in using the Arecibo telescope, the 
constraints of telescope time availability, and the principal science objectives 
outlined in Section \ref{science}. Here we review the considerations
that enter into the survey design and numerical  simulations that have been used to
refine it.

\subsection{Scaling Relations \label{scaling}}

The HI mass of an optically thin HI source at distance $D_{Mpc}$, in solar units, is
\be 
M_{HI}/M_\odot = 2.356\times 10^5 D^2_{Mpc} \int S(V) dV,
\ee
where $S(V)$ is the HI line profile in Jy and $V$ is the Doppler velocity
in \kms. To first order,
\be
M_{HI}/M_\odot \simeq 2.4\times 10^5 D^2_{Mpc} S_{peak} W_{kms},
\ee
where $S_{peak}$ is the line peak flux and $W_{kms}$ its velocity
width in \kms. For detection, the signal--to--noise ratio
$s = f_\beta S_{peak}/S_{noise}$ must exceed some threshold value;  
$f_\beta\leq 1$ quantifies the fraction of the source 
flux detected by the telescope's beam. The parameter $f_\beta =1$ for a point 
source, while for resolved sources, it decreases roughly like the ratio 
between the beam solid angle and the solid angle subtended by the source. 
An estimate of $S_{noise}$ can be obtained from 
the radiometer equation for the rms figure
\be
S_{rms} = {(T_{sys}/G) \over \sqrt {2\times \Delta f_{ch}\times t_s\times f_t}},
\ee
where $T_{sys}/G$ is the system temperature divided by the system 
gain (for the ALFA feeds, $T_{sys}/G$ will vary between 2.65 and 3.40 Jy; 
here we adopt a flat value of 3.25 Jy); $\Delta f_{ch}$ is the channel bandwidth in 
Hz and $t_s$ the integration time in seconds. 
The factor 2 under the square root indicates that two independent 
polarization channels are averaged. For ALFALFA, $\Delta f_{ch} = 25$ kHz, 
which at the rest frequency of the HI line, is equivalent to 5.3 \kms. The 
factor $f_t$ accounts for spectral smoothing of the signal, $f_{smo}$, 
the switching technique applied for bandpass subtraction, $f_{switch}$, 
and other observational details, such as autocorrelation clipping
losses, i.e. $f_t = f_{switch}f_{smo}f_{other}$. For the data taking 
scheme of ALFALFA, $f_{switch}f_{other} \simeq 0.7$.
The signal--to--noise of a feature of width $W_{kms}$ is best rendered
when the noise is measured after smoothing the signal to a spectral resolution
on order of $W_{kms}/2$. In practice, however, the smoothing of L--band
spectra of $W_{kms}\simeq$ several 
hundred \kms does not reduce the noise in proportion to $W_{kms}^{1/2}$ 
and, moreover, $S_{peak}$ is depressed by such smoothing, for spectral 
shapes are by no means boxlike. The fact that the detection criterion
described above applies well to narrow lines but not so to wider ones
was also noted by Rosenberg \& Schneider (2002). We assume here that spectral
smoothing will increase signal--to--noise up to a maximum $W_{kms}\simeq 200$,
and that smoothing beyond that width will be ineffective in increasing $s$.
For a conservative signal--to--noise threshold of 6, we can then write:
\be
12.3 f_\beta t_s^{1/2}\Bigl({M_{HI}\over 10^6 M_\odot}\Bigr) D^{-2}_{Mpc} 
\Bigl({W_{kms}\over 200}\Bigr)^\gamma > 6,  \label{eq:sngt5}
\ee
where $\gamma=-1/2$ for $W_{kms}<200$ and $\gamma=-1$ for $W_{kms}\geq 200$.
By inverting, we can obtain a minimum detectable HI Mass
\be
\bigl({M_{HI}\over 10^6}\bigr)_{min} = 0.49 f_\beta^{-1} D_{Mpc}^2
~t_s^{-1/2} (W_{kms}/200)^{-\gamma}.
\label{eq:MHI}
\ee
With an integration time of 30 sec per pixel solid angle (see Section 
\ref{sensitivity}) or 48 sec per beam solid angle,  
ALFALFA should thus detect an HI mass of $10^6$ $M_\odot$, $W_{kms}=25$,
at a distance of $\sim 6.5$ Mpc, and a source of $10^7$ $M_\odot$ and of 
the same width out to $\sim 20$ Mpc.

It is useful to review some of the basic scaling relations relevant to the design 
of a survey:

\noindent $\bullet$ The minimum integration time required to detect a 
source of HI mass $M_{HI}$ and width $W_{kms}$ at $s=6$, at the distance $D_{Mpc}$ 
with ALFA is, from eqn. \ref{eq:MHI},
\be
t_s \simeq 0.023 f_\beta^{-2} \Bigl({T_{sys}\over G}\Bigr)^2 \Bigl({M_{HI}\over 10^6 
M_\odot}\Bigr)^{-2} (D_{Mpc})^4 
\Bigl({W_{kms}\over 200}\Bigr)^{-2\gamma}, \label{eq:ts}
\ee
i.e. {\bf the depth of a survey increases only as $t_s^{1/4}$}.
With equality of back--ends, the $t_s$ required to detect
a given $M_{HI}$ at a given distance decreases as the square of $G$,
i.e. as the 4th power of the reflector diameter. Arecibo offers a
tremendous advantage because of its huge collecting area.

\noindent $\bullet$  
The beam of a telescope of collecting area $A$ is $\Omega_b \propto A^{-1}$, 
while the maximum distance at which a given HI mass can be detected is 
$D_{max} \propto G^{1/2}$. Since $G\propto A$, 
the volume sampled by one beam to the maximum distance $D_{max}$ is 
$V_{beam}\propto \Omega_b D_{max}^3/3 \propto A^{1/2}$,
i.e. in a fixed time, a radio telescope samples a 
{\bf volume that scales with the reflector diameter}, yielding
a significant comparative advantage for a large aperture like Arecibo.

\noindent $\bullet$  Assuming that clouds of mass $M_{HI}$ are randomly 
distributed in space out to the maximum distance at which they are 
detectable, $D_{max}(M_{HI})$, the number of clouds detected by a 
survey increases linearly with the sampled volume 
$V_{survey} = \Omega_{survey} D_{max}^3/3$, where $\Omega_{survey}$ is 
the solid angle mapped by the survey. 
We can thus increase the number of detections either by sampling a 
larger solid angle $\Omega_{survey}$ or by increasing $D_{max}(M_{HI})$. 
Now, the total time required to complete the survey is 
\be t_{survey} \propto (\Omega_{survey}/\Omega_b) t_s, \label{eq:tsurvey}
\ee
where $\Omega_b$ is the telescope beam. 
Since $D_{max}(M_{HI}) \propto t_s^{1/4}$, as shown in equation
\ref{eq:ts}, we can write
\be
V_{survey}(M_{HI}) \propto \Omega_{survey} [D_{max}(M_{HI})]^3
                    \propto \Omega_{survey} t_s^{3/4} \propto t_{survey} t_s^{-1/4},
\ee
and inverting:
\be
t_{survey} \propto V_{survey}(M_{HI}) D_{max}(M_{HI})
           \propto V_{survey}(M_{HI}) t_s^{1/4}.
\label{eq:tsurvey2}
\ee
To achieve a given surveyed volume $V_{survey}(M_{HI})$, once $M_{HI}$ is
detectable at an astrophysically interesting distance, {\bf it is 
more advantageous to maximize $\Omega$ than to increase the
depth of the survey $D_{max}(M_{HI})$}.

The scaling relations described above provide only general guidelines
in the design of a survey. Other considerations can and will play 
important roles in the survey strategy. For example, the growing
impact of RFI on HI spectroscopy dictates increased attention to
signal identification and corroboration, recommending a survey with
more than a single pass over a given region of sky, as we discuss in
Section \ref{2pass}. The determination
of specific properties of galaxies or systems may drive towards deeper
surveys of narrow solid angle regions, as planned for other ALFA
surveys with the Arecibo telescope, the goals and products of which
will be complementary to ALFALFA.

\subsection {Survey Simulations \label{simul}}

The scaling relations described above dictate that ALFALFA cover a
very large solid angle. In practice, the survey design must weigh the
desire to cover a wide area with the need for sensitivity.
An indispensable aid in the design of a survey is a thorough examination
of expectations, {\it vis--a--vis} variance over the survey parameter
space. To this end, we have carried out an extensive set of survey 
simulations to help in the design of ALFALFA and present
a sample of the results in this section.

The main ingredients for our survey simulation are: (i) the survey mode and sensitivity
parameters, deriving from the instrument configuration; (ii) an estimate of 
the space density of sources given by an adopted  HIMF; (iii) an understanding of the clustering
properties and deviations from smooth Hubble flow in the local Universe. 
Sensitivity considerations were presented in Section \ref{scaling}. For the HIMF, 
we use two recent estimates which differ strongly from each other at the low
mass end: that of Z97 and that of RS02. The more recent HIMFs by Z04 
and Springob \etal (2005a) are bracketed by those of Z97 and RS02.
We use a density map of the local Universe provided by Branchini \etal (1999),
which is a density reconstruction derived from the PSCz catalog. The grid we used has a
spacing of 0.9375 $h^{-1}$ Mpc in the inner 60 $h^{-1}$ Mpc, and a
spacing twice that value between 60 and 120 $h^{-1}$ Mpc, where $h=0.7$;
the map is smoothed with a Gaussian filter of $\sigma=3.2 h^{-1}$ Mpc. For distances
larger than 120 $h^{-1}$ Mpc, we assume a constant density. 

Sources are seeded using the density map and, separately, each of the two HIMFs.
The HI gas is assumed to be optically thin. Because many of the sources will
be resolved by the Arecibo beam, an estimate of HI sizes is necessary. Assuming
that the HI distribution is disk--like, inclinations to the line of sight and
linewidths need to be assigned to each source. We use empirical scaling relations
obtained from our own HI survey data (Springob \etal 2005b) and Broeils \& Rhee (1997),
we add random inclinations, realistic scatter and broad--band spectral baseline 
instability. With these recipes, we have inspected a wide grid of
survey parameters in arriving at the adopted ALFALFA survey strategy.

\begin{figure}[h]
\hskip -3cm
\plotone{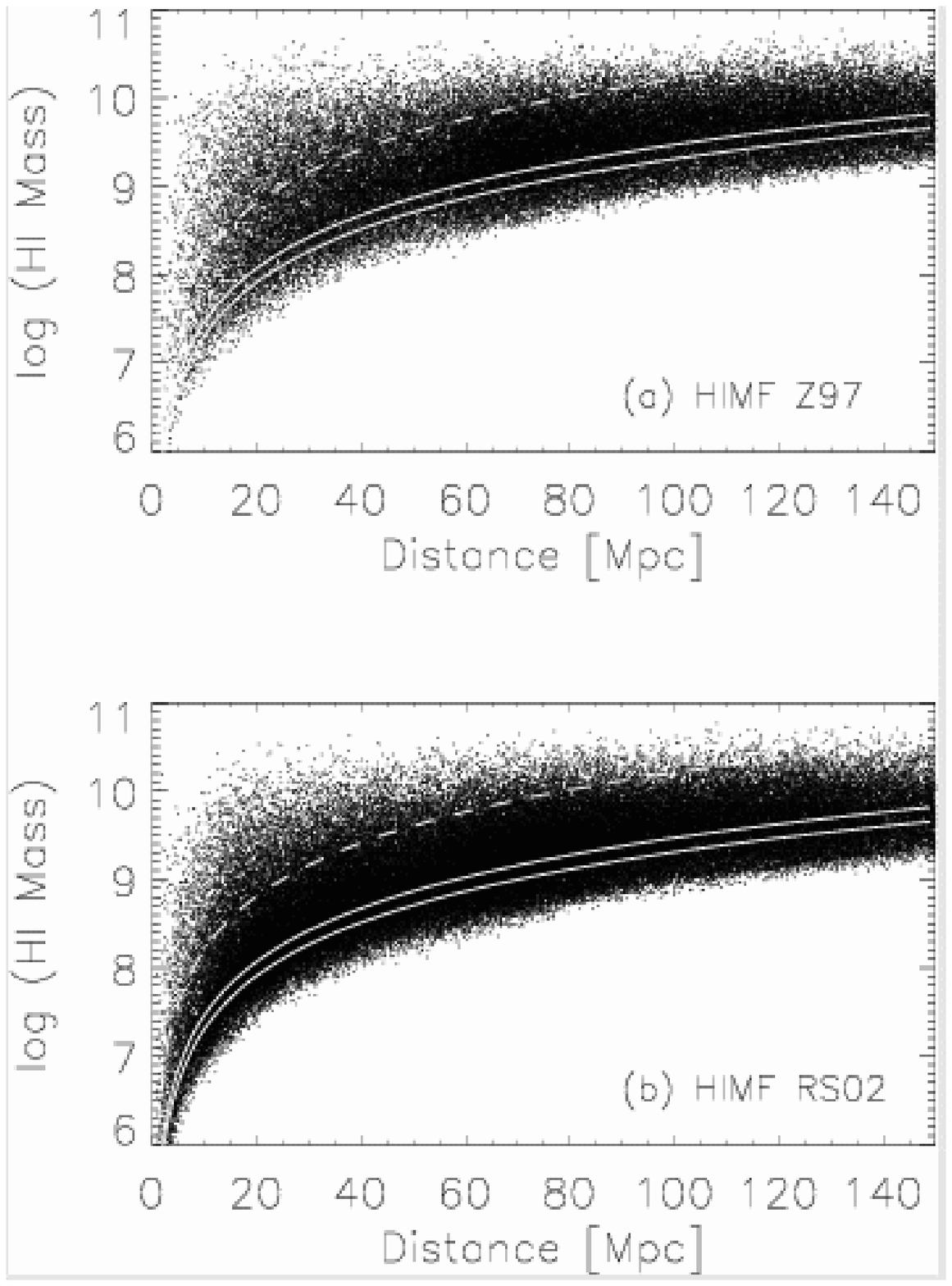}
\vskip -4cm
\caption{HI Mass vs. distance plot of expected detections within $D<150$ Mpc,
assuming a Z97 HIMF (panel a) and a RS02 HIMF (panel b). Calculations were
made for the ALFALFA sky region specified in Section \ref{tiles}, $t_s=30$ 
seconds per map pixel area and a detection threshold $S/N=6$. The solid 
lines in both panels indicate detection
limits of 0.9 Jy \kms and 1.25 Jy \kms. The first would be near
the completeness limit of the survey for sources of width $< 200$ \kms,
the second near the completeness limit for objects of the same width, but
with $t_s=14$ seconds integration. The dashed line corresponds to a
flux density of 6.8 Jy \kms, a 6--sigma HIPASS detection limit for a 200 
\kms wide source.}
\label{HIM_D}
\end{figure}

As an example, Figure \ref{HIM_D} displays the variation with distance 
of the HI Mass of expected detections within the region of the ALFALFA sky coverage, as 
described in Section \ref{tiles}, adopting as input the two different
HIMFs: the Z97 HIMF (panel a) and the RS02
(panel b). Only expected detections out to 150 Mpc are shown. 
The simulation corresponds to an
integration time $t_s=30$ seconds per map pixel solid angle. The number of
detections expected with RS02 is 22,200, while the number expected with
Z97 is 15,022, with a detection threshold of $S/N=6$, as defined in Section 
\ref{scaling}. The difference in expected detections is more dramatic when
the HI mass of the source is restricted to \mhi ~$<10^8$ \msun. In
that case, we expect 1400 detections with RS02 and only 249 with Z97.
It is interesting to point out that in the course of ALFALFA precursor
observations, reported in the companion Paper II, three objects with
\mhi ~$<10^7$ \msun ~were detected. Albeit of still marginal statistical 
value, that rate is consistent with the high end of the expectations 
(RS02 HIMF) obtained from the simulations reported here.

Three curves are inset in the panels of Figure \ref{HIM_D}: the
two solid lines are the loci of constant integrated HI line flux of 0.9 and 1.25 
Jy \kms, respectively. The lowest of the two corresponds to the completeness 
limit of the survey for sources of $W_{kms}\leq 200$ \kms (detections
below that line correspond to sources of smaller width). The second curve
corresponds to completeness limit for sources of the same width, for an
integration time per pixel of $t_s =14$ seconds. Such an integration time
applies to the analysis of individual drift tracks, without the corroborating
support (and higher resulting integration) of spectra in beam tracks at
neighboring Declinations. This detection limit would result if source
extraction would be carried out, for example, right after data taking,
and before an entire data cube (spatially two--dimensional, plus one 
spectral dimension) is available. In this case, the expected number 
of detections would be 13,804 for the RS02 HIMF, and 9601 for the Z97
case, a drop of respectively 38\% and 36\% from the previous set of
numbers. The decrease in the number 
of detections with small HI masses would be more severe if signal 
extraction were applied to individual tracks only, rather than to full
maps: in that case, only fewer than half of the sources would be 
detected in the RS02 case, and just above half in the Z97 case.

The topmost (dashed) curve inset in  Figure \ref{HIM_D} corresponds
to a flux integral of 6.8 Jy \kms, the HIPASS completeness limit at
the 6$\sigma$ level, for detection of sources of width $\leq 200$ \kms;
this is the HIPASS analog of the lowest of the two solid lines for ALFALFA.
It uses a HIPASS limit of 13.3 mJy per map pixel area, as reported
by Barnes \etal (2001). This provides a good graphical illustration
of the comparison between the two surveys.

\begin{figure}[h]
\epsscale{1.2}
\hskip -5cm
\plotone{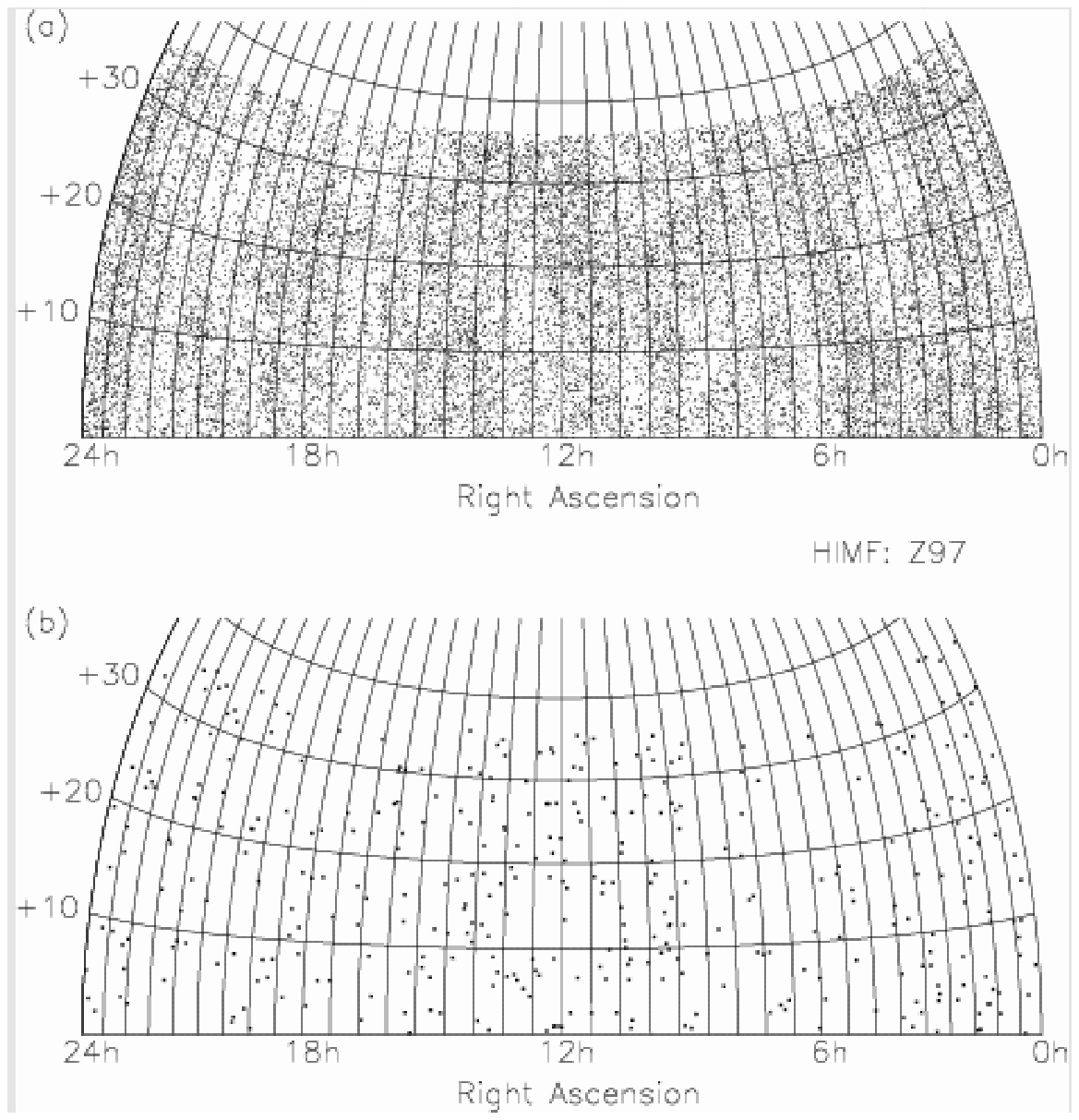}
\vskip -15cm
\caption{Sky distribution of expected detections for the ALFALFA survey,
as rendered with a Z97 HIMF. In panel (a) sources of all HI masses are
plotted, while in panel (b) only those with $M_{HI} < 10^8$ $M_\odot$
are shown. Note that the ALFALFA survey will be restricted to Right Ascensions
$07^h$ to $16.5^h$ and $22^h$ to $3^h$, although the full range of R.A.s
is shown in the figure.}
\label{aitoffa}
\end{figure}

\begin{figure}[h]
\epsscale{1.2}
\hskip -5cm
\plotone{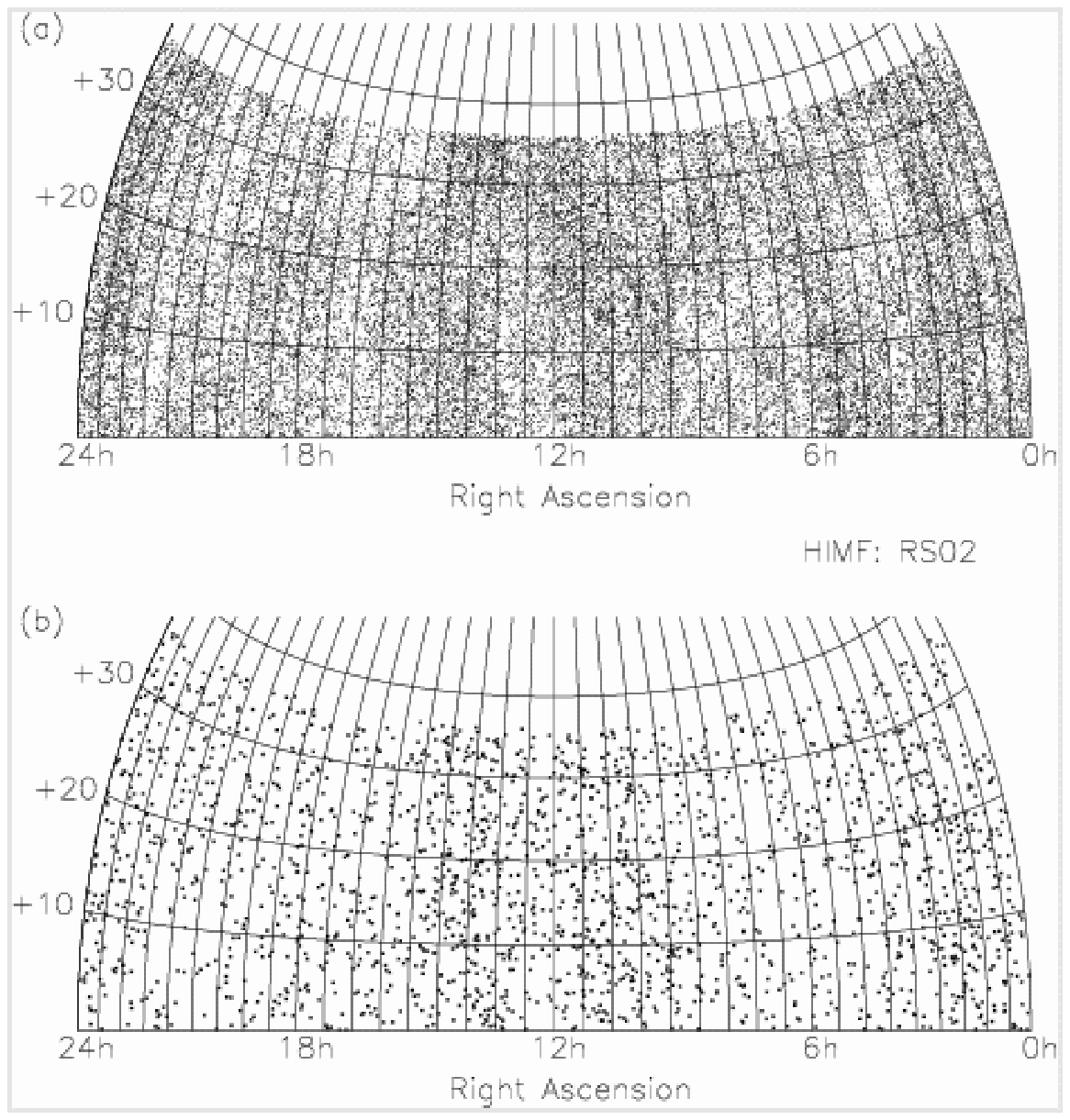}
\vskip -15cm
\caption{Analogous display to that in Figure \ref{aitoffa}, except that
the RS02 HIMF was used for the simulation.}
\label{aitoffb}
\end{figure}

Figure \ref{aitoffa} shows the sky distribution of the detected sources
by an ALFALFA--like survey, in the simulation with the Z97 HIMF. Sources
at all Right Ascensions are plotted, albeit ALFALFA will only cover 60\%
of the full R.A. range. In the lower panel, only the detections with
\mhi ~$<10^8$ \msun ~are plotted. Figure \ref{aitoffb} shows the 
analogous graphs for the RS02 HIMF.

Simulations results such as those presented here aid us in the estimate of 
the statistical efficacy of the survey data, most importantly in the determination
of the faint end of the HIMS and the clustering properties of the new detections.

\section{Observing Mode}\label{obs}

Given the science objectives outlined in Section \ref{science} and the
scaling relations and simulations presented in the preceding section,
the final consideration of the survey design strategy takes into
account more telescope-related practicalities. In this section, we
review those issues which have led us to adopt a very simple observing
strategy, a two-pass drift scan mode, covering the sky with ``tiles''
extending from  $0^\circ < {\rm Dec.} < +36^\circ$.

\subsection{Drift Mode}\label{drift}

The Arecibo telescope is an altitude--azimuth system located at a latitude
near 18$^\circ$. Its Gregorian dome can be
steered within $\sim 20^\circ$ of the zenith, but the system gain and
performance degrades rapidly at zenith angles above $\sim18^\circ$. In general,
the performance characteristics including beamwidth, pointing accuracy,
sidelobe levels, spectral baseline stability and susceptibility to RFI vary 
with both azimuth, zenith angle and feed rotation angle. Furthermore, the 
ALFA footprint on the sky and its beams' structure vary likewise in a 
complicated manner. The design 
of ALFA surveys is thus strongly constrained by this variance, and with it, the 
corresponding degree of calibration complexity a particular observing program 
can endure. ALFALFA aims to minimize the impact of these factors on performance
through a choice of maximum simplicity in the observing mode and minimal
electronic intrusion at the detection level.

As mentioned previously in Section \ref{alfa}, the ALFALFA survey is
thus being carried out in a fixed azimuth drift mode. For most of the 
survey, the azimuth arm of the telescope is stationed along the local 
meridian, the zenith angle of ALFA determining the declination to be 
mapped. A tiny elevation readjustment is periodically applied to 
maintain drift tracks at constant J2000.0, rather 
than current declination. Without such adjustment, drift tracks taken
few years apart would noticeably diverge from one another. No firing of
noise calibration diodes is done during normal data acquisition. Rather,  
data taking is interrupted very briefly every 10 minutes while a
calibration noise diode is fired for 1 second. This interruption
produces data gaps of 5 seconds --- approximately 1/3 of a beamwidth
in R.A. --- in each 600 second drift data stream. The central
frequency of the bandpass is set and never changed during an observing
session, i.e. no Doppler tracking of the local oscillator frequency is 
applied (realignment to a common heliocentric reference frame is applied 
to each spectrum off line, using high precision time, position and 
Earth's motion stamps updated every second in data headers).
With no moving telescope parts, constant gain and nearly constant
system temperature along a drift are 
obtained; standing waves will change slowly, as driven by the sidereal rate; 
beam characteristics remain fixed; bandpass subtraction is optimized.

The solid angle
mapped by the survey is subdivided into ``tiles'' of $4^\circ$ in 
Declination (see below and Figure \ref{skycov}) extending from Dec.=$0^\circ$ 
to $36^\circ$. For 8 of the 9 bands of tiles, the azimuth of the
feed array is along the local meridian (at azimuth either $0^\circ$ or 
$180^\circ$) while the rotation angle of the 
feed array is fixed at 19$^\circ$, yielding tracks
for the seven array beams that are equally spaced in Declination.  
In order to map the telescope's ``zone of avoidance'' near zenith,
the band of tiles centered at Dec.=$+18^\circ$ will require a 
different orientation: with the azimuth arm nearly E-W. 
This strategy greatly simplifies the disentangling of main beam
and sidelobe contributions to the maps: characterization of ALFA 
parameters thus needs to be made on a greatly reduced volume of telescope 
configuration parameter space. 

Drift mode observations, combined with the calibration scheme described above,
yield maximally efficient use of telescope time, providing high photometric
quality with very small overhead. We expect that, bar instrumental
malfunctions, telescope time usage for science data will approach 97\%.

\subsection{Two--Pass Strategy \label{2pass}} 

As discussed in Section \ref{scaling}, the volume sampled at any HI mass 
limit, for a survey of fixed total duration, varies with the integration 
time per point as $t_s^{-1/4}$. Once a threshold sensitivity 
is reached, it is more advantageous to increase the solid angle of the 
survey than its depth. Because of the spacing of the ALFA beam tracks
in drift mode discussed in Section \ref{alfa}, coverage of the sky 
in a one--pass drift survey is slightly worse than Nyquist. For a
fixed amount of observing time, a single pass strategy would appear to
maximize the number of detected sources. For a fixed total survey
time, the loss of survey volume sampled by going from a one--pass to a 
two--pass drift survey is, according to Section \ref{scaling}, 19\%. 
Several advantages of a two--pass strategy offset that loss, however: 
(1) A second pass will greatly aid the separation of cosmic emission from
RFI which is unlikely to affect each pass identically; (2) The 
denser sky sampling will allow statistical separation of spurious signals
from cosmic ones, thus allowing reliable detection to lower values of
S/N; (3) If the two passes are made when the Earth is at very different 
phases of its orbit, confirmation of cosmic nature for detection candidates 
can be obtained by verifying that they are separated in topocentric radial 
velocity by $30\cos(\Delta \theta)$ \kms, where $\Delta\theta$ is the change 
in the angle between the line of sight to the detection candidate and the 
velocity vector of Earth on its heliocentric orbit; this requires that 
the second pass be undertaken 3 to 9 months after the first pass, modulo
a year; (4) The variability of radio continuum sources can
be measured, and radio transients can be identified, allowing
commensality with other science teams interested in studies of those
phenomena; (5) Given design features of the ALFA hardware,
maintenance will be difficult and, as a 
result, ALFA may operate at less than 100\% capacity (i.e. one or more
beams may be unusable) during some fraction of the time.
Loss of a beam in a single--pass survey would result in grievous holes
in sky coverage, whereas 
a two--pass strategy would greatly attenuate the resulting damage to the survey.
For all these reasons, the high galactic latitude Arecibo sky to be
mapped by ALFALFA will be covered in two drift passes. The resulting effective
integration time of the survey, per beam area, will be about 48 seconds.
Another way of expressing the sensitivity of the survey is in terms of
the integration time per \sqd, which will be about 14,700 seconds.

\subsection{Sky Tiling and Data Products\label{tiles}} 

As shown in Figure \ref{skycov}, the sky to be mapped by ALFALFA extends 
between $0^\circ < {\rm Dec.} < 36^\circ$ and over two blocks of Right 
Ascension, respectively $07^h 30^m$ to $16^h 30^m$
and $22^h 00^m$ to $03^h 00^m$, although the vagaries of telescope time 
allocation will produce some irregularities in the survey solid angle boundaries.
The exclusion of the low galactic latitude regions within the telescope's
horizon is driven by (a) the realistic assessment that pulsar and other 
galactic ALFA surveys will greatly increase the pressure on low galactic
latitude LSTs and (b) the expectation that part of the low galactic
latitude, extragalactic sky will be surveyed commensally with pulsar and
other galactic surveys. 

For bookkeeping and data release purposes, the sky mapped by ALFALFA will be 
subdivided into 378 {\it tiles}, each of $20^m$ in R.A. and $4^\circ$ in Dec.
Mapping a tile in single--pass drift mode requires 17 drifts of ALFA, spaced 
$\sim14$\arcmin ~in Dec. and each yielding 7 drift tracks; equally as many 
additional drifts 
are required to complete the second pass at a later time. For the second pass,
beam tracks will be interleaved with those of the first pass, so that the 
final Declination sampling will be $\sim1'$, better than Nyquist. In
order to minimize ``scalloping'' of the gain over the map 
introduced by the higher gain of central beam relative to the outer
ones, the second pass drifts are offset by 7\arcmin18\arcsec
~relative to the first pass tracks.

The data processing environment chosen for ALFALFA is IDL. A substantial
body of spectral line software generated by one of us (PP) already exists
at the Arecibo Observatory. Further development specific to ALFALFA has 
been grafted on this fertile base. The tile size was chosen to 
constitute a data block that can reasonably be handled for data processing 
in an efficient manner by current desktop computers. The generation of raw 
data proceeds at the rate of $\sim 1.2$ GB/hr, and upon conversion from
its raw FITS format to an IDL structure, a single 600 sec drift 
is $\sim 200$ MB. Such a data block is well suited for 
one of the most computer intensive parts of the reduction pipeline, that of 
bandpass subtraction. The data for a full tile, after polarization averaging 
and regridding, can fit within the 2--4 GB memory of current low--cost
desktops. 

The data processing path for ALFALFA
data can be summarized as follows:
\begin{itemize}
\item One FITS file per 600--record drift is generated by the data taking
software at the Arecibo Observatory. By the end of each observing session,
each of those is converted into an IDL structure, a `drift' structure, and 
stored for further analysis at the Observatory and the observers' institutions.
\item Within weeks, all data of an observing session is noise--calibrated and
a bandpass solution is computed. The ``bandpassed'', calibrated and baselined
spectral data for each beam/polarization configuration are obtained as 
output of an automated pipeline that is designed to preserve not only small 
angular scale features such as external galaxies, but also large structures 
such as HVCs and galactic HI. 
\item The first detailed visual inspection of the data follows, in the
course of which the observer flags regions of each position--velocity map
for RFI and other occurrences of data corruption. It is anticipated that
as much as about half of all sources to be detected by ALFALFA will be
visible to the eye at this stage. A first automated signal extraction
algorithm pass will produce a list of candidate detections. Noise 
diode--calibrated, bandpass--corrected, baselined and RFI--flagged spectra 
as obtained to this stage constitute what we shall refer to as {\it Level 
I Data Products}. 
\item Upon completion of the second pass through a given sky tile, data
will be re--calibrated using the continuum sources present within the tile,
regridding of the sky sampling will take place, after smoothing by a 
homogenoeus resolution kernel and conversion into data cubes will follow.
The output of this processing stage shall be referred to as {\it Level II Data 
Products}.
\end{itemize}
Because telescope scheduling is a dynamic process which responds to proposal
pressure at a national center, the scheduling of data releases far in advance 
is not possible. However, the ALFALFA observing status is continuously
updated at the survey website\footnote{\it  http://egg.astro.cornell.edu/alfalfa}. 
Observing plans foresee completion by bands of tiles and, when an accurate
prediction of the completion is available, data release plans for that
band will be posted. Data release will take place through an ALFALFA/HI node 
connected to the U.S. National Virtual Observatory.
A preliminary example of web--based data presentation is linked to the
aforementioned website as well as directly reachable
\footnote{\it http://egg.astro.cornell.edu/precursor}: it allows access to 
the spectral line data, images of optical counterparts and parameter tabulation 
of the precursor observations' results discussed in Paper II.

\begin{figure}[h]
\epsscale{1.2}
\hskip -5cm
\plotone{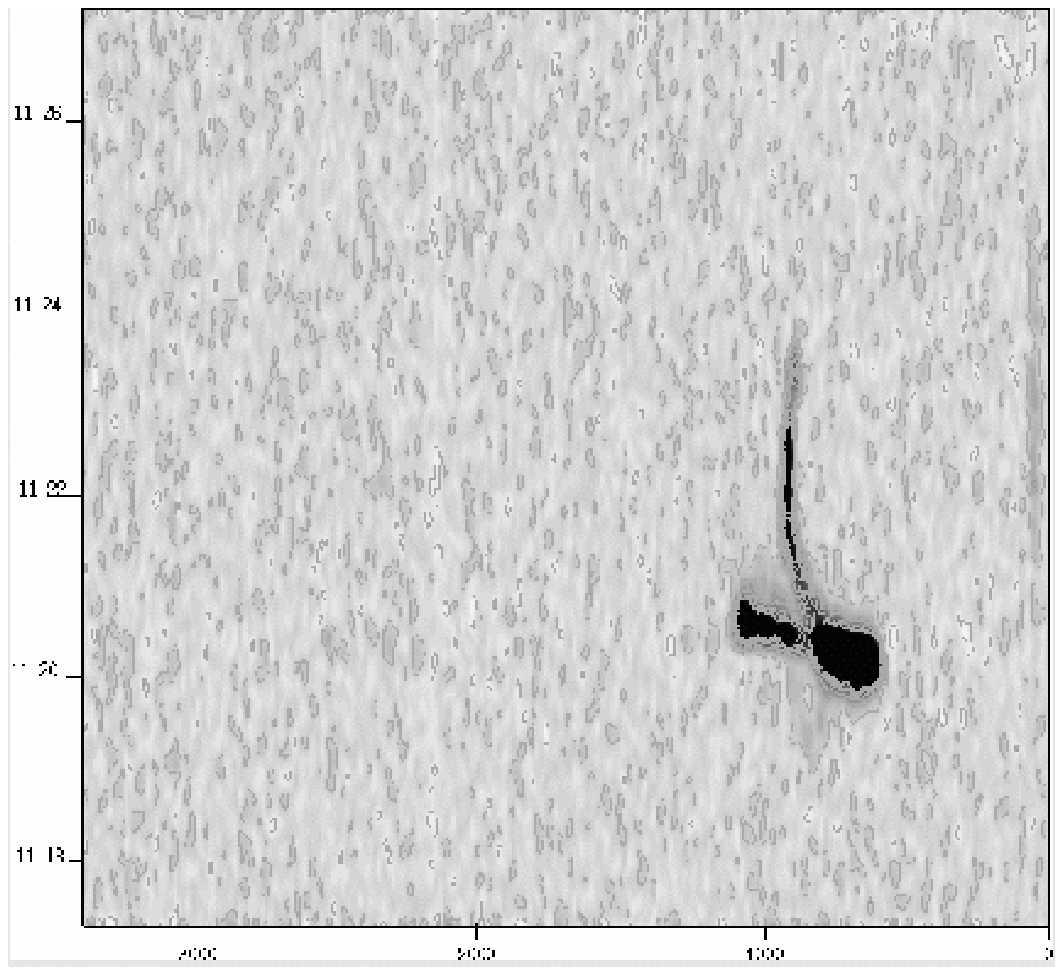}
\vskip -20cm
\caption{Position-velocity map cutting across NGC3628 and its tidal
tail at constant Dec(J2000) = $+13^\circ$ 36\arcmin 45\arcsec. The lowest contour
is at 2 mJy per beam, and other contours are spaced by 6 mJy per beam. See
text in Section \ref{tiles} ~for further details.
\label{n3628}}
\end{figure}

As an example of a Level I data product and the potential of the
ALFALFA survey, Figure \ref{n3628} shows a single pass
drift across the galaxy NGC~3628, one of the Leo Triplet galaxies, and its tidal tail. 
The displayed position--velocity 
image consists of a constant declination drift of 600 1-sec records at
Dec. (J2000) = $+13^\circ$36\arcmin 45\arcsec, corresponding to 
Dec. (B1950) = $+13^\circ$53\arcmin 09\arcsec ~for comparison with Figure 1 of Haynes, 
Giovanelli \& Roberts (1979). The single drift rms noise in the image is 3.5 mJy. 
Contours are linearly spaced by 6 mJy and the lowest contour is plotted at
2 mJy per beam. The tidal tail is traced by the ALFALFA data
as far as the earlier point-by-point map, but a countertail at earlier Right Ascension than
NGC 3628 is also clearly visible in the new map. After the second pass
ALFALFA data become available, the Leo Triplet region will be mapped with a sensitivity
twice as deep as that shown in Figure \ref{n3628}, allowing a detailed
study of the system over a wide area.

\section{Expected Survey Sensitivity}\label{sensitivity}

Here we summarize the sensitivity of ALFALFA at several survey levels:
\begin {itemize}
\item A 1--second record of a drift scan, after accumulation of both polarizations,
will yield a spectrum of $S_{rms}\simeq 13(res/10)^{-1/2}$ mJy, where
$res$ is the spectral resolution in \kms.
\item A single drift, position--frequency map spatially smoothed to the spatial 
resolution of the telescope beam will yield $S_{rms}\simeq 3.5(res/10)^{-1/2}$ 
mJy.
\item A spatially two--dimensional map of two--pass ALFALFA data, smoothed with 
a kernel of 2' at half power, will have $S_{rms}\simeq 2.3(res/10)^{-1/2}$ 
mJy per pixel.
\item The rms sensitivity per beam area, after a two--pass survey, will be
$S_{rms}\simeq 1.8(res/10)^{-1/2}$.
\item The 6$\sigma$ HI column density limit will be
$N_{HI,lim}=1.6\times 10^{18} (W/10) (res/10)^{-1/2}$ atoms cm$^{-2}$, 
for a spectral line of width $W$ \kms, observed with a spectral resolution 
of $res$ \kms.
\end{itemize}

\begin{figure}[h]
\epsscale{0.8}
\plotone{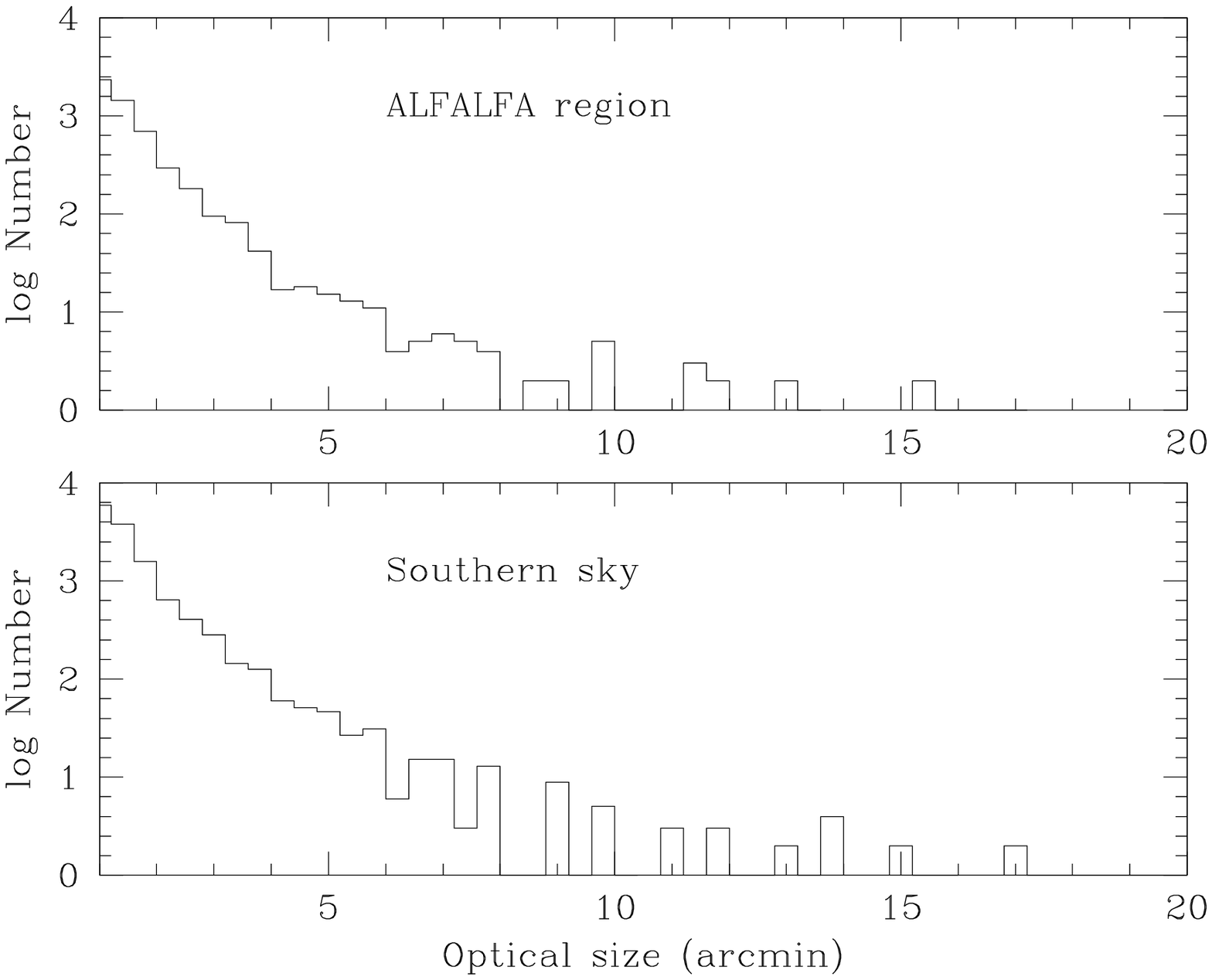}
\caption{Histograms of the optical major blue diameter, D25, of
galaxies larger than 1\arcmin, in the 
ALFALFA survey region (upper) and the whole southern
hemisphere. Aperture synthesis studies have shown that the diameter
of the HI disk for optically selected galaxies is on average
1.6 times larger than the
optical size. The Parkes 15\arcmin ~beam thus resolves on order of 15 galaxies
over the whole southern hemisphere; the ALFA $\sim$3.5\arcmin ~beam should resolve several
hundred galaxies over the ALFALFA survey region. The bin size is 0.1\arcmin ~in
both histograms.}
\label{sizehis}
\end{figure}

Column density sensitivity is, in general, independent of telescope size, and
thus ALFALFA will not reach deeper $N_{HI}$ levels than previous wide angle
surveys such as HIPASS. In fact, given the shorter integration time per beam
area, ALFALFA will have lower sensitivity to $N_{HI}$ than did HIPASS
for very extended sources. It may
be argued that only surveys with longer integration times per beam area than
HIPASS can break new ground. However, this argument holds true only if
sources are well resolved by the telescope beam. The beam area of
Arecibo is nearly {\it 20 times smaller} than that of the Parkes telescope.
If sources are unresolved
by the beam, the telescope can only detect total flux, and the observation
cannot be used for any inference on source column density. In fact,
very few extragalactic HI sources were resolved by the Parkes beam;
the smaller Arecibo beam size gives ALFALFA a major advantage. To
illustrate this point, Figure \ref{sizehis} shows two histograms of
the angular size distribution of optically selected, catalogued
galaxies: the upper one for galaxies known to be within the ALFALFA survey
region, the lower one for galaxies in the whole of the southern hemisphere.
The optical size used for this comparison is D$_{25}$ as catalogued in
the Third Revised Catalog of Galaxies (de Vaucouleurs \etal ~1991). It
is well established that, on average, the HI size (measured at the level near
1 M$_\odot$ pc$^{-2}$) of optically selected galaxies is about 1.6 times that 
blue size (Broeils \& Rhee 1997), although for dwarf irregular systems that
ratio may rise significantly (Swaters \etal 2002). Even allowing for
a small number of extremely large HI-to-optical size ratios, the total number of galaxies 
resolved by the Parkes telescope beam over the whole of the southern sky is 
on the order of a dozen or two. Only for those, the Magellanic Stream and High Velocity
Clouds is column density sensitivity of any relevance for HIPASS. The ALFALFA survey
should, on the other hand, resolve several hundred galaxies and High Velocity
Clouds, and map their peripheries to a column density limit of the order
of $5\times 10^{18}$ atoms cm$^{-2}$. Through careful analysis of the
HI mapping datasets including consideration of the impact of
sidelobe contamination in the two or three selected telescope
configurations adopted for the drift survey, the ALFALFA survey will address the
issue of whether a column density regime below $10^{19}$ cm$^{-2}$ is
commonly found in the local Universe (Corbelli \& Bandiera 2002).

\section{Candidate Detections and Verification of Cosmic Signals}\label{followup}

ALFALFA will produce a catalog of tens of thousands of candidate
detections; on order of 20,000 will be cosmic sources. The number of 
candidate detections per bin of signal--to--noise $s$ will increase 
steeply, and the probability that the candidates are real sources
decreases rapidly with diminishing $s$ value. 

Internal (i.e. within the survey data set) corroboration of 
candidate detections will rely on (a) comparison of independent
polarization samples and (b) comparison of spatially adjacent
survey samples. The effectiveness of part (b) will depend on the 
spatial sampling density and, in the case of multiple drifts through
the same region, on the temporal consistency of the data. These
comparisons will help exclude many marginal candidate detections
of non--cosmic origin, which we shall refer to as ``false'' candidates.

Post--survey, corroborating observations will be desirable to 
confirm candidate detections just below the signal--to--noise threshold
above which signal corroboration can be internally possible. This may
allow significant expansion of the survey `catch' with modest additional
amounts of telescope time. The usual compromise is necessary in setting a 
$s$ threshold: too high a threshold will lose many valuable potential 
detections; too low a threshold will require impractical amounts of 
post--survey telescope time; a haphazard criterion may corrupt the 
completeness of acquired samples.

We expect that the bulk of follow--up observations to corroborate
the cosmic nature of detection candidates will be carried out at
Arecibo, using a single--pixel feed, hopping from one candidate
to the next, minimizing slew and setup time. Two modes may lead to
low efficiency usage of telescope time: too dense a set of
follow--up targets may be comprised of too large a fraction of false
candidates, and thus produce a low yield per unit of telescope time; 
too sparse a set may lead to a large fraction of the time spent 
slewing and in other overhead. Careful optimization will be required. 
We consider some of these issues in this section.

\subsection{Types and Numbers of Candidate Detections}

Visual inspection or automated signal extraction algorithms identify 
detection candidates that can be assigned to three classes: (i) cosmic 
sources, (ii) extreme statistical fluctuations of noise and (iii) 
spurious signals due to RFI or other instrumental and data analysis 
causes.

\begin{figure}[h]
\plotone{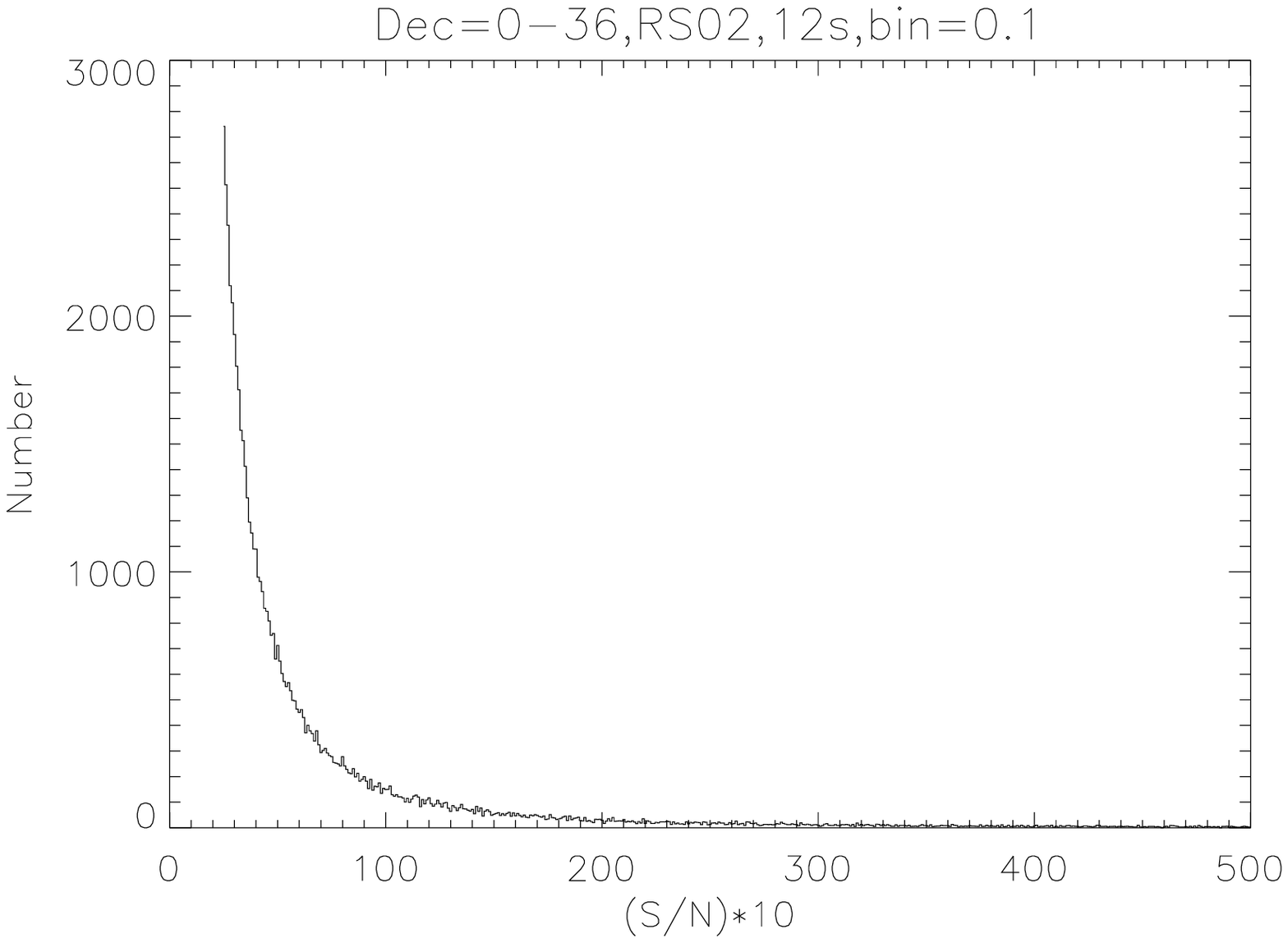}
\caption{S/N histogram of a 12s ALFALFA survey using the RS02 HIM.
S/N bins have width 0.1 in S/N. S/N is defined as the peak signal flux to
the rms, computed in matched filter mode, over a spectral resolution
equal to 1/2 the signal width. Only tentative detections with
S/N$>2.5$ are plotted. \label{stnhisto}}
\end{figure}

\noindent {\it Cosmic Sources.} Figure \ref{stnhisto} shows a histogram
of signal--to--noise $s$ of expected sources, as obtained in one of the 
simulations described in Section \ref{simul}. We emphasize that the number 
of sources rises steeply as $s$ decreases, illustrating the well--known 
fact that most of the survey candidate detections will occur near the 
detection limit.

Because of the massive bulk of ALFALFA data sets, signal extraction
will largely rely on automated procedures. Amelie Saintonge has coded a
matched--filter, cross--correlation signal extraction algorithm,
described elsewhere (Saintonge \etal, in preparation). Detection probability 
simulations have been carried out 
with this algorithm, by randomly injecting a Gaussian signal in simulated 
spectra with Gaussian noise, and monitoring the effectiveness of the signal 
extraction algorithm in recovering the injected signal. The detection
probability is computed as the fraction of all trials in which the 
signal extraction algorithm positively identifies the injected signal;
such probability is monitored as a function of $s$ and of signal width. When
the noise is measured after spectrally averaging
over 1/2 the spectral width of the injected signal, the detection 
probability is largely independent of the signal width. 


\noindent {\it Statistical noise fluctuations.} For Gaussian noise,
the probability that a single spectral channel yield a fluctuation
of signal--to--noise between $s_1$ and $s_1+ds_1$ is
\be
p_1 ds_1 = {1\over \sqrt{2\pi}} e^{-s_1^2/2} ds_1,
\ee
where $s_1=S_{peak}/\sigma_1$, with $S_{peak}$ the peak flux density and 
$\sigma_1$ the rms noise, with single--channel spectral resolution.
Similarly, the probability for a $n_w$ channels--wide spectral
feature to exhibit a deviation between $s_n$ and $s_n+ds_n$ is
\be
p_n ds_n = {1\over \sqrt{2\pi}} e^{-s_n^2/2} ds_n,
\ee
where $s_n = s_1/\sqrt{n_w/2}$.

In a survey of $N_{los}$ line of sight samples, taken with a
spectrometer of $N_{c}$ channels, the number of samples 
$n_w$ channels wide, with signal--to--noise between $s$ and $s+ds$ is
\be
n_{s,n_w} ds = N_{los} {N_c\over n_w} p_n ds = 
N_{los} {N_c\over n_w} {1\over \sqrt{2\pi}} e^{-s^2/2} ds
\ee
and the total number of statistical fluctuations of that width
with $s$ larger than a threshold $s_{th}$ is
\be
N_{s_{th},1} = N_{los} {N_c\over n_w} {1\over \sqrt{2\pi}} 
\int_{s_{th}}^{\infty} e^{-s^2/2} ds = N_{los} {N_c\over n_w} 
[F(\infty)-F(s)],
\ee
where, again, the noise is computed with a spectral 
resolution of $n_w/2$ channels, and $F(s)$ is the familiar
cumulative distribution of the normal error function: $F(-\infty)=0$,
$F(\infty)=1$ and $F(0)=0.5$. The total number of purely statistical
noise fluctuations between $s_a$ and $s_b$, with widths between
$n_{w1}$ and $n_{w2}$, appearing in the survey will then be
\be
N_{[a,b],[1,2]} = \sum_{n_w=n_{w1},n_{w2}} N_{los} {N_c\over n_w} 
[F(s_a)-F(s_b)] .
\ee

To first order, and ignoring the fact that in the expression above
a high $s$, broad feature gets overcounted as several, lower $s$, 
narrower ones, we can approximate
\be
N_{[a,b],[1,2]} \simeq N_{los} N_c ~\ln {n_{w2}\over n_{w1}}~[F(s_a)-F(s_b)].
\label{Nab12}
\ee

For example, for a survey that samples $10^7$ lines of sight, with 
a spectrometer usefully covering 85 MHz with $N_c=3600$ spectral 
channels (so that a velocity width range between 25 and 500 \kms
translates into $n_{w1}\simeq 5$ and $n_{w2}\simeq 100$, one should
expect $N_{>3}\sim 2\times 
10^8$ features with $s>3$, $N_{>4}=5\times 10^6$ features with 
$s>4$, and $N_{>5}=5\times 10^4$ features with $s>5$, with width 
anywhere between 25 and 500 \kms. These are ominously large numbers 
when compared to expected numbers of cosmic sources between $10^4$
and $3\times 10^4$, for that range of $s$.

\noindent {\it RFI or other Spurious Signals.} The above assumption 
of Gaussian noise is heuristic. The true nature of the noise
will be ascertained after a significant fraction of the survey
data will have been collected and the ``normal'' characteristics
of both equipment and RFI environment will have been measured.
As we discuss in Paper II, the precursor observations carried out
in 2004 were taken in commissioning mode for the ALFA hardware,
and several internal bugs have been found and fixed since those
observations were completed. Those observations are not well suited
for a careful analysis of the problem. For the moment, we will ignore 
the impact of RFI and other non--Gaussian sources of noise, and
restrict our analysis to the discrimination between cosmic
sources and Gaussian noise fluctuations.

\subsection{Discriminating Among Candidate Detections}\label{vicinity}

Assuming that the majority of weak detection candidates will
be unresolved by the telescope beam, the most important means
of discriminating between cosmic sources and noise fluctuations
will result from comparison of contiguous drift tracks and 
different polarizations of the same beam. 
Consider a single--pass ALFALFA survey, whereby contiguous 
drift tracks are separated by 2.1\arcmin ~in Declination on the sky. 
A point source swept by one of the feeds will appear also in 
contiguous tracks, at lower $s$. For an ALFA beam averaging 
3.5\arcmin ~width at half power, the response 2.1\arcmin ~off the beam center 
is about 0.37 of that on beam center. In a two--pass drift
survey, the track separations are 1.05\arcmin, and at that distance
from the beam center, the beam response is 0.78 of that on
beam center. A point source will thus be far more easily confirmed
in a two--pass survey. Similarly, since the HI line is unpolarized,
the comparison of the two independent polarization spectra of the 
same beam will deliver equal signals, plus noise, for cosmic
sources, and completely uncorrelated results for noise.
We shall refer to the exclusion of detection candidates made
possible by comparison of adjacent drift tracks and polarization
channels as {\it vicinity trimming}, and distinguish three sets of
detection candidates: (1) that obtained without any vicinity
trimming, and those (smaller ones) obtained after (2) vicinity trimming in a
single--pass survey and (3) vicinity trimming in a double
pass survey.

\begin{figure}[h]
\plotone{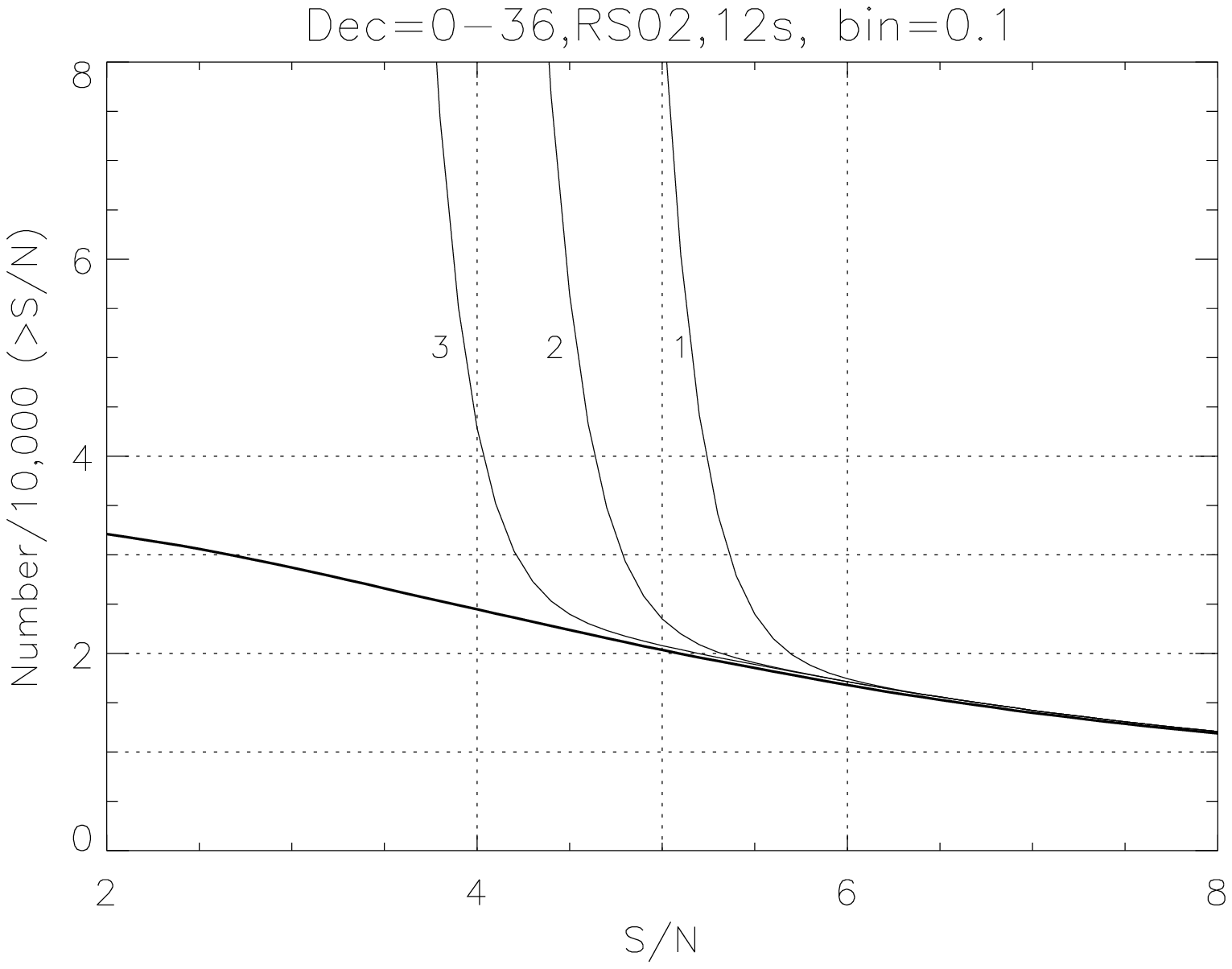}
\caption{Cumulative number of candidate detections as a function of S/N, 
expected for an all--Arecibo sky drift survey. The lower (nearly
flat) line refers to the 'real' sources (assuming an RS02 HIMF)
while the upper curves refer to the three cases of `vicinity
trimming' described in Section \ref{vicinity}. \label{cumdets}}
\end{figure}

Figure \ref{cumdets} ~displays the cumulative number of
candidate detections plotted as a function of signal--to--noise,
expected for an ALFALFA--like survey. The simulation assumes
a RS02 HIMF and a sensitivity corresponding to a double--pass drift 
survey. The thick line corresponds to the detection candidates
associated with cosmic (`real') sources. The three thin lines
correspond to the expected number of noise fluctuations with the 
given $s$ in the three `vicinity trimming' cases described above.
At $s>6$, most of the detection candidates are cosmic sources.
At $s\simeq 5$, the number of `real' cosmic sources is 25\%
higher than at $s=6$, but the candidate detections resulting
from noise fluctuations is several times higher
than that of real sources. Vicinity trimming can however 
drastically reduce the number of candidate detections deserving
attention. Only at signal--to--noise levels $s<4$ does the
number of noise fluctuations overwhelm that of cosmic sources,
after two--pass vicinity trimming. At this level, however,
the impact of low power level RFI will play an important
and yet quantitatively unknown role.

\subsection{Follow--up, Corroborating Observations}

Candidate detections with a comfortably high signal--to--noise
threshold $s_{th}$ will not need corroborating follow--up
observations in order to confirm their reality as cosmic sources.
Without considering the impact of RFI, $s_{th}$ may be in the
vicinity of 6; consideration of the impact of RFI may
raise $s_{th}$ to higher values, in a variable manner depending
on the frequency of the candidate signal. Candidates with
$s\simeq s_{th}$ or slightly below that level will be reobserved
with the Arecibo telescope. To how low a level of $s$ should
re--observations be considered? A simple criterion would be
that corroborating observations should be requested only for
candidates of $s$ such that the expectation of confirmation,
expressed in terms of detections per unit of telescope time,
is at least as high as for the full survey. 

If a corroborating observation is to require an increase in
the $s$ from a value of, say, 5 for the survey data, to about
7 for the corroborating observation, integration times of at least
1 minute per candidate will be necessary for corroborating
observations. Observing runs to corroborate several hundred candidate 
detections at a time will thus be the norm. If the set of candidate
detections to be checked is very sparse --- say one candidate every several
square degrees --- slew times will be very substantial and 
bandpass--correcting observations will be required for each
candidate source, more than doubling the required telescope time. 
In that case, the on--source $t_{int}$ of order of 1 minute 
may be a small fraction of the overall time 
required to observe each source. The Arecibo telescope slew times
are respectively $0.4^\circ$ s$^{-1}$ in azimuth and $0.04^\circ$ 
s$^{-1}$ in elevation. A 1$^\circ$ change in elevation will
require 25 seconds. It will thus be observationally advantageous
if the sky density of tentative sources to be corroborated is
high, e.g. on order of one per square degree or higher. Not only
will that reduce the overhead of slew motions and settle time, 
but it will also allow for a running mean bandpass to be accumulated 
over a few contiguous staring observations, as the telescope 
configuration would change little between adjacent source 
candidates. In that case, allowing for slew and settle time, a 
corroborating observation of a single source will require on 
order of one to two minutes of telescope time. The steeply
rising fraction of `false' sources with decreasing $s$
suggests that corroborating observations requesting single
pixel telescope time at the level of approximately 10\% 
of the request proposed for the ALFA observations will
deliver optimal returns.

\section{Summary} \label{summary}

ALFALFA uses the new 7-beam Arecibo L-band feed array to carry out
a wide area survey of the high galactic latitude sky visible from
Arecibo. In addition to the all--important sensitivity advantage that
accrues from using Arecibo, the world's most sensitive radio telescope
at L--band, ALFA offers
important and significant improvements in angular and spectral resolution
over the available major wide area extragalactic HI line surveys such
as HIPASS and HIJASS. ALFALFA is intended to produce an extensive database of HI spectra
that will be of use to a broad community of investigators, including many interested in
the correlative mining of multiwavelength datasets. It is specifically designed to probe
the faint end of the HIMF in the very local Universe.

As a result of practical considerations and simulations of survey
efficiency, ALFALFA exploits a simple fixed-azimuth drift scanning --- 
{\it minimum intrusion} --- technique. A two--pass strategy will
greatly aid in the rejection of spurious signals and RFI, thus minimizing the need
for follow-up confirmation observations, evening out the scalloping in the maps that
arises from unequal pixel gain, and offering the opportunity to use the same dataset
for the statistical characterization of continuum transients. 
Initial tests of the hardware, software and survey observing mode,
conducted in Fall 2004 during the ALFA commissioning phase as
described in Paper II confirm the efficacy of the planned approach.
The basic parameters of the ALFALFA survey can be summarized 
accordingly as follows:
\begin{itemize}
\item Sky coverage of 7074 \sqd, between $0^\circ$ and $+36^\circ$
in Declination, 7.5 to 16.5 and 22.0 to 3.0 hrs in Right Ascension, with
3.5\arcmin ~spatial resolution.
\item Frequency coverage between 1335 and 1435 MHz, yielding coverage
of extragalactic HI in redshift out to $cz<18,000$ \kms,
with 5.3 \kms maximum spectral resolution.
\item Sensitivity of $1.8\times(res/10)^{-1/2}$ mJy per beam area, where
$res$ is the spectral resolution in \kms.
\item On order of 20,000 HI sources are expected to be detected by the
survey. Extragalactic HI sources with 
\mhi ~$\simeq 10^6$ \msun ~will be detectable to a distance of 6.5
Mpc, while HI masses \mhi ~$\simeq 10^7$ \msun, will be detectable
throughout most of the Local Supercluster, including the Virgo cluster
and out to 20 Mpc. Several hundreds will have \mhi ~$< 10^{7.5}$
\msun, thus allowing a robust determination of the faint end of the HIMF.
\item Public access data products will be produced on a continuing
basis as subsets (tiles) of the overall survey are completed.
\end{itemize}

Observations for ALFALFA started in February 2005, and completion of the
survey is expected to take five to six years. Cataloguing a complete
census of HI locally, pinning down the HIMF to the
lowest masses and conducting the blind HI absorption and OH megamaser surveys
will require completion of the full 5-year program, but even the
initial 2005 allocation promises
early science results in several important areas including: the mapping of 
nearly 1600 \sqd, more than 3 times the coverage with
twice the sensitivity of the Arecibo Dual Beam Survey (Rosenberg \& Schneider 2000);
a first blind census across the Virgo cluster region with a detection limit
of M$_{HI} > 10^7$ \msun ~at the cluster distance (assuming a width W = 30 \kms);
a complete search for HVCs around M33;
the identification of gas-rich galaxies in the NGC~784 and Leo~I groups;
the mapping of the environments of 12 gas-rich galaxies with D$_{UGC}
>$ 7\arcmin; and a first attempt at a large blind survey for HI absorbers.
 
\vskip 0.3in
 
RG and MPH acknowledge the partial support of NAIC as Visiting Scientists during
the period of this work. This work has been supported by NSF grants AST--0307661,
AST--0435697, AST--0347929, AST--0407011, AST--0302049; DGICT of Spain grant
AYA2003--07468--C03--01;  
and by a Brinson Foundation grant. We thank the Director of
NAIC, Robert Brown, for stimulating the development of major ALFA surveys, Enzo 
Branchini for providing a PSCz density grid in digital form,
H\' ector Hern\' andez for a sensible 
and friendly approach to telescope scheduling, the Director, telescope 
operators and support staff of the Arecibo Observatory for their proactive 
assistance.

\vfill
\end{document}